\documentclass[12pt,preprint]{aastex}

\newcommand{\DD}[2]{\frac{d #1}{d #2}}
\newcommand{\DP}[2]{\frac{\partial #1}{\partial #2}}
\newcommand{\DDt}[2]{\frac{d^{2} #1}{d #2{}^{2}}}
\newcommand{\DPt}[2]{\frac{\partial^{2} #1}{\partial #2{}^{2}}}

\slugcomment{}

\shorttitle{}
\shortauthors{Ishitsu and Sekiya}

\begin{document}


\title{Stabilization of the Shear Instability 
in a Dust Layer of a Protoplanetary Disk 
and Possible Formation of Planetesimals 
due to Gravitational Fragmentation of the Dust Layer}


\author{Naoki Ishitsu}
\affil{Astronomical Data Analysis Center,
National Astronomical Observatory,
Osawa 2-21-1, Mitaka, Tokyo 181-8588, Japan}
\email{naoki.ishitsu@nao.ac.jp}

\and
\author{Minoru Sekiya}
\affil{Department of Earth and Planetary Sciences, Faculty of Sciences,
33 Kyushu University, Hakozaki, Fukuoka, 812-8581, Japan }



\begin{abstract}
We show that the planetesimal formation due to the gravitational fragmentation of a dust layer in a protoplanetary disk is possible.  The dust density distribution in the dust layer would approach the constant Richardson number distribution due to the dust stirring by the shear instability and dust settling. We perform the analysis of the shear instability of dust layer in a protoplanetary disk with the constant Richardson number density distribution. Our study revealed that this distribution is stable against the shear instability even if the dust density at the midplane reaches the critical density of the gravitational instability, and the planetesimal formation through the gravitational fragmentation of the dust layer can occur even for the dust to gas surface density ratio with the solar composition. 

\end{abstract}


\keywords{planetary systems: protoplanetary disks---solar system: formation---hydrodynamics---instabilities
}


\section{Introduction}

It is in controversy whether planetesimals are formed in a protoplanetary disk by the gravitational instability \citep{safronov69, goldreich73, coradini81, sekiya83} or by collisional sticking of dust aggregates due to some non-gravitational forces \citep{weiden84, cuzzi93, wurm01}. 

The planetesimal formation through dust sticking due to non-gravitational forces has some difficulties. 
A meter-sized body falls to the inner edge of the disk due to the gas drag in much shorter time scale ($\sim 10^3$ yr) \citep{adachi76, weiden77} than the possible life time of a disk ($\sim 10^7$ yr). 
\citet{cuzzi93} 
have suggested that a body sweeps smaller dust aggregates as the body falls towards the central star and can grow to form a km-sized planetesimal. 
Moreover, \citet{wurm01} have observed the growth of a body due to the following mechanism:  When a dust aggregate impacts a target body, the aggregate fragments into monomers. 
After the monomers bound on the body several times, they finally stick to the body.
However, this experiment was done with a condition in which the target body 
was smaller than the gas mean free path. 
\citet{sekiya03} 
have indicated that the dust 
%MS% aggregates 
aggregate %MS%
breaks into monomers by the first collision and the monomers do not collide the target body again if the body is larger than the gas mean free path, because the monomers are carried away by the gas flow. 
Thus, the body can no longer grow by sweeping monomers if the body grows up to a radius on the order of the mean free path of the gas. 
For example, the body can not grow larger than about 1cm at 1AU. 

On the other hand, the formation of planetesimals due to the gravitational instability is attractive since the particular sticking mechanism of dust aggregates is not needed. 
Moreover, by growing from millimeter-sized dust aggregates to a 
km-sized body within a short period, the process avoids the problem that meter-sized bodies fall towards the central star by the gas drag. 
However, it has been considered that the gravitational instability of the dust layer is difficult to occur if a disk is turbulent; the dust aggregates are stirred up from the midplane, so that the dust layer can not reach the critical density above 
which the gravitational instability of the dust layer arises. 

The local turbulence would occur even though the disk is laminar, that is, the global turbulence such as the thermal convection does not exist. \citet{weiden80} has suggested that the turbulence is induced due to the vertical shear in the dust layer. 
The turbulence diffuses the dust aggregates and prevents them from settling towards the midplane.  
The cause of the vertical shear is explained as follows.
The gas tends to revolve with a sub-Kepler velocity, balancing the gravitation of the central star, the centrifugal force, and the gas pressure gradient. On the other hand, dust tends to revolve with the Kepler velocity, balancing the two former forces.  Thus, supposing the coupling between dust and gas is good, their velocity is determined by
\begin{equation}
	v_{\phi} = \left[1- (\eta \rho_g/ \rho)\right]v_K, 
	\label{eqn:unv}
\end{equation} 
where 
\begin{equation}
\rho = \rho_g + \rho_d,
\label{eqn:rho}
\end{equation}		
\begin{equation}
 \eta = - r \left( \partial P /\partial r \right)/(2 v_K^2 \rho_g ), 
	\label{eqn:eta}
\end{equation}		
where $\rho_g$ is the spatial gas density, $\rho_d$ is the spatial dust density
(i.e. the total mass of dust in unit volume of the protoplanetary disk),
$v_K$ is the Kepler velocity, $r$ is the distance from the rotation axis of the disk, 
and 
$P$ is the gas pressure. Thus shear occurs in the dust layer, depending on the dust to gas density ratio \citep{weiden80, cuzzi93, sekiya98}. 

\citet{cuzzi93} and \citet{sekiya98} have argued that the 
dust diffusion by shear-induced turbulence and the dust settling by the vertical component of the gravity of a central star lead to an 
equilibrium dust distribution. The strength of the shear is characterized by the Richardson number $J$, which is defined by 
\begin{equation}
	J = - g_z (1/\rho) (\partial \rho / \partial z)
	 (\partial v_{\phi} /\partial z)^{-2}, 
\end{equation}
where $z$ is the distance from the midplane, and $g_z$ is the gravitational acceleration in the $z$-direction.
In a shear flow with density stratification and the gravitational force, but without the effects of the Coriolis and the tidal forces, the Richardson number $J$ is used in the criterion for stability. 
The physical meaning of $J$ is a quarter of the ratio of the potential energy consumed by the perturbation to the extra energy gained by the perturbation from the unperturbed shear flow.
If $J>J_c=1/4$, more energy is consumed than is gained by the perturbation;
such a motion is inhibited by the energy conservation law and thus the flow is stable. 
On the other hand, the flow can be unstable below the critical Richardson number $J_c$. 
\citet{sekiya98} derived the equilibrium state as follows: 
As dust aggregates settle towards the midplane, 
the local Richardson number decreases below the critical value $J_c$ in a certain region of the dust 
layer. Then the turbulence is induced by the shear instability, which 
would stir dust aggregates. 
As the dust aggregates diffuse, the shear rate $|\partial v_{\phi}/\partial z|$ decreases, the local Richardson number increases, 
and the shear-induced turbulence 
ceases after $J$ exceeds the critical value $J_c$. 
Then the dust settling starts again. 
Repeat of the processes is expected to bring to the constant Richardson number dust density distribution (CRNDDD in the followings) with $J=J_c$ for the entire region of the dust layer if dust aggregates are sufficiently small and 
are stirred up by very weak turbulence. 

The dust density can not reach the critical density of the gravitational instability for CRNDDD with $J=0.25$ if the dust to gas surface density ratio is that calculated from the solar abundance. 
In contrast, the gravitational instability could occur when the dust to gas surface density ratio increases, that is, the dust concentrates or the gas depletes by some mechanisms \citep{sekiya98, youdin02, youdin03}.  

In order to elucidate the shear instability in more detail, we have performed a series of researches.  
\citet{sekiya00} have performed the linear analysis of the shear instability for CRNDDD.  
Their results have shown that the growth rate of the shear instability is less than the Kepler angular frequency. 
Next we investigated the instability using a different unperturbed state \citep{sekiya01}.
As dust particles stick each other and settle towards the midplane, the density distribution in the dust layer tends to be constant \citep{watanabe00}.
Thus, we employed the hybrid dust density distribution (called HDDD in the following) in which the dust layer has a constant density and transition regions have sinusoidal density distributions. 
For this distribution, the growth rate of the shear instability is much larger than the Kepler angular frequency when the dust density at the midplane is larger than the gas density; the latter is much less than the critical density of the gravitational instability. 

However these works did not include the Coriolis force and the tidal force. 
\citet{ishitsu02} have investigated the stability of HDDD including the effect of the Coriolis force but neglecting the tidal force. 
Their results have shown that the Coriolis force has little effect 
on the growth rate of the shear instability for HDDD. 
 However, the mode of the instability changes from the co-rotation resonance to resonances like the Lindblad resonance. 

 \citet{ishitsu03} have investigated the shear instability of HDDD including both the tidal force and the Coriolis force. 
That is, they have employed the dust distribution expected before the onset of the shear instability. 
The paper indicated that the tidal force stabilizes the shear instability if the growth rate of the shear instability in the case without the tidal force is comparable to or less than the Kepler angular velocity. 
However, the shear instability occurs before the dust density reaches the critical density of the gravitational instability. 
Thus, turbulence due to the shear instability would change the dust density distribution. 
As a result, the dust distribution is expected to approach CRNDDD asymptotically.
In this work, we investigate the linear stability of CRNDDD, including the Coriolis and tidal forces and the self-gravity of the dust layer. 

In \S 2, the constant Richardson number distribution is introduced. In \S 3, the basic equations for the linear analysis are derived. In \S 4, results of the case with the Coriolis force only are shown. In \S 5, results of the case with the Coriolis and tidal forces are shown. The results of the case with the self-gravity of the dust layer are also shown. 
In \S 6, possibility of the planetesimal formation due to the gravitational instability is discussed.

\section{The Constant Richardson Number Dust Density Distribution}
Here, the constant Richardson number dust density distribution (CRNDDD) 
derived by \citet{sekiya98} is reviewed.
As written in the previous section, \citet{sekiya98} considered that the dust density distribution asymptotically approaches 
CRNDDD with $J=J_c$, 
and he analytically derived 
it (note that his solution is also applicable to the case without $J=J_c$; 
thus we here assume only that $J=$constant): 
\begin{equation}
\frac{|z|}{\sqrt{J} \eta r} = \sqrt{u^2-u_m^2}-q 
\ln\left[ \left(u+\sqrt{u^2-u_m^2} \right)/u_m\right],
\label{eqn:Richard}
\end{equation}
where
\begin{equation}
	u=\rho_g/[\rho_d(z) + \rho_g] + q,
\label{eqn:u}
\end{equation}
and 
\begin{equation}
	q = 4 \pi G \rho_g/ \Omega_K^2. 
\end{equation}
 Here, 
$G$ is the gravitational constant,
$\Omega_K$ is the Kepler angular velocity, and 
$u_m$ is the value of $u$ at the midplane $z=0$.
The assumption that the gas density $\rho_g$ is constant within the dust layer 
is appropriate if the thickness of the dust layer is much thinner than 
the gas scale height $H_g$. Then gas density is given by
\begin{equation}
	\rho_g = \Sigma_g / (\sqrt{\pi} H_g), 
\end{equation}
where the disk is assumed isothermal in the $z$-direction and $\Sigma_g$ is the gas surface density: 
\begin{equation}
	\Sigma_g = 1.7 \times 10^{3} f_g (r/\mbox{AU})^{-3/2} \mbox{[g cm${}^{-2}$]},
\end{equation}
where $f_g$ is the ratio of our gas surface density to that of the Hayashi model \citep{hayashi81, hayashi85}.
The dust surface density is given by 
\begin{eqnarray}
\Sigma_d &=& 2 \int^{z_d}_{0} \rho_d dz  \nonumber\\
   &=& 2 \sqrt{J} \eta r \rho_g \left\{ (1+q) 
            \ln \left[ \left(1+q + \sqrt{(1+q)^2 -u_m^2} \right)/u_m \right] 
	-\sqrt{(1+q)^2 -u_m^2} \right\} \nonumber\\ 
   &=& 7.1 f_d (r/\mbox{AU})^{-3/2} \mbox{[g cm${}^{-2}$]},
   \label{eqn:sigmad} 
\end{eqnarray}
where the half of thickness of the dust layer $z_d$ is given from equation (\ref{eqn:Richard}) by
\begin{equation}
	z_d/(\sqrt{J} \eta r ) = \sqrt{(1+q)^2 -u_m^2} - q \ln 
	\left[ \left(1+q + \sqrt{
	(1+q)^2 - u_m^2 } \right) /u_m \right],
\end{equation}
and $f_d$ is the ratio of the dust surface density to the Hayashi model for the region where the water vaporizes. 
In the disk with the gas to dust mass ratio of the solar abundance, 
\begin{eqnarray}
 	f_d/f_g = \left\{
\begin{array}{ll}
 4.2  & \mbox{for the region where water condenses} \\
 1    &	\mbox{for the region where water vaporizes}.
\end{array}
\right.
\end{eqnarray} 
.

\citet{sekiya98} considered that CRNDDD with $J=J_c$ is realized as written in \S 1.  
He showed that the midplane density for CRNDDD with $J=J_c$ is much lower than the critical density of the gravitational instability. 
He also showed that the the midplane density increases if $f_d/f_g$ increases by dust concentration and/or gas depletion by some processes.  
Note here that the critical density of the gravitational instability used by 
\citet{sekiya98} $\rho_d(0)/\rho_g=260f_g^{-1}(r/\mbox{AU})^{-1/4}$ is that for the constant dust density distribution \citep{sekiya83}, although 
\citet{sekiya98} investigated the stability of CRNDDD.
Recently, 
\citet{yamoto04} 
calculated the critical density of the gravitational instability for 
CRNDDD: 
$\rho_d(0)/\rho_g=750f_g^{-1}(r/\mbox{AU})^{-1/4}$. 
The explanation for this difference is that the effective thickness of the dust layer for CRNDDD
is thinner than the thickness for the constant dust density distribution 
if the dust and gas surface density are same.
Thus the CRNDDD requires the larger critical density for the gravitational instability.
Using this new value, the dust density at the midplane reaches the critical density if the dust surface density is enhanced by $f_d=17.15$, 
or the gas surface density is reduced by  $f_g=0.0286$
 compared to Hayashi's dust surface density 
at 1AU \citep{hayashi81, hayashi85}. 
These values are not so different from those obtained by 
\citet{sekiya98}, $f_d=16.8$ and $f_g=0.029$, 
because the midplane density increases rapidly 
as $f_d$ increases or $f_g$ decreases \citep{yamoto04}.

\section{Basic Equations}
This section gives the basic equations under some assumptions described below; they are similar to those written in \citet{ishitsu03}, but include the self-gravity. 
We neglect the curvature of the cylindrical coordinates $(r,\phi, z)$ and use the local Cartesian coordinate system which rotates with the Kepler velocity 
$v_K(R)$ at $r=R$ where $R=$constant 
\citep{goldreich78}. 
Our coordinates $x, y$ and $z$ denote the radial, 
the azimuthal and the vertical 
directions of the disk, respectively. 
That is, $x=r-R$, $y=R[\phi-\Omega_K(R)t]$, and $z=z$, 
where $\Omega_K(R)$ is the Kepler angular velocity at $r=R$ 
and we neglect higher order terms of $x$, $y$ and $z$. 
In the following, we denote 
$v_K(R)$ 
and $\Omega_K(R)$ by $v_K$ and $\Omega_K$, respectively, 
for simplicity. 
In the Keplerian disk, the flow experiences the tidal force $3 \Omega_K^2 x$. 
The gas can be assumed incompressible because the dust layer treated here is much thinner than vertical scale height of the gas disk.
The mixture of gas and dust is considered as one fluid because 
we treat dust aggregates whose frictional times are much shorter than the Kepler period and the growth time of the shear instability. 
This assumption is fairly good if dust aggregates are smaller than e.g. 1cm at 1 AU. 
Thus, we obtain the continuity equation, the mass conservation equation, the momentum equations, and the Poisson's equation
 \begin{equation}
	  \DP{u}{x} + \DP{v}{y} + \DP{w}{z} =0,
	 \label{eqn:ba1}
 \end{equation}
 \begin{equation}
	 \DP{\rho}{t} +u \DP{\rho}{x} + 
	 v\DP{\rho}{y} + w\DP{\rho}{z} =0,
	 \label{eqn:ba2}
 \end{equation}
\begin{equation}
	 \DP{u}{t} + u \DP{u}{x} + v\DP{u}{y} + w\DP{u}{z} 
      = - \frac{1}{\rho} \DP{P}{x} + 3 CT \Omega_{K}^2 x +2 C \Omega_{K}v 
	-\DP{\Psi}{x},
	\label{eqn:ba3}
\end{equation} 
 \begin{equation}
	  \DP{v}{t} + u \DP{v}{x} + v \DP{v}{y} + w\DP{v}{z} 
       = - \frac{1}{\rho} \DP{P}{y}  -2 C \Omega_{K} u -\DP{\Psi}{y},
	 \label{eqn:ba4}
 \end{equation}
 \begin{equation}
	  \DP{w}{t} + u \DP{w}{x} + 
	 v \DP{w}{y} +w \DP{w}{z} 
       = - \frac{1}{\rho} \DP{P}{z} -  \Omega_K ^2 z -\DP{\Psi}{z},
	 \label{eqn:ba5}
 \end{equation}
 \begin{equation}
	   \DPt{\Psi}{x} + \DPt{\Psi}{y} + \DPt{\Psi}{z} 
       =  4 \pi G \rho,
	 \label{eqn:ba6}
 \end{equation}
where $u, v$ and  $w$ are the radial, the azimuthal and the vertical velocities, respectively, and $\Psi$ is the gravitational potential. 
If $C=1$ and $T=0$, the Coriolis force is taken into account, but 
the tidal force is not taken into account.
On the other hand, if $C=T=0$, 
neither the Coriolis force nor the the tidal force are taken into account.
Of course, $T=C=1$ for a real flow, but we used these parameters to investigate the effects of the tidal and the Coriolis forces by comparing the results with $T=C=0$, $T=0$ and $C=1$, and $T=C=1$. 
For $u=P=\Psi=0$, equation (\ref{eqn:ba3}) reads 
\begin{equation}
v= - \frac{3}{2}  T \Omega_K  x.
\label{eqn:keplerv} 
\end{equation}
This expresses the circular Kepler motion in the local coordinate system.
 In order to eliminate the Keplerian part of the velocity, we introduce the velocity relative to the Keplerian motion,
\begin{equation}
\bar{v} = v + \frac{3}{2} T \Omega_K x.
\end{equation} 
From equations (\ref{eqn:ba1}) to (\ref{eqn:ba6}), we have
\begin{equation}
	 \DP{u}{x} + \DP{\bar{v}}{y} + \DP{w}{z} =0,
	\label{eqn:bta1}
\end{equation}
\begin{equation}
	\DP{\rho}{t} +u \DP{\rho}{x} + 
	(\bar{v} - \frac{3}{2}T \Omega_K x)\DP{\rho}{y} + w\DP{\rho}{z} =0,
	\label{eqn:bta2}
\end{equation}
\begin{equation}
	 \DP{u}{t} + u \DP{u}{x} + 
   (\bar{v} - \frac{3}{2} T\Omega_K x) \DP{u}{y} + w\DP{u}{z} 
      = - \frac{1}{\rho} \DP{P}{x} + 2 C \Omega_{K}\bar{v} -\DP{\Psi}{x},
	\label{eqn:bta3}
\end{equation}
\begin{equation}
	 \DP{\bar{v}}{t} + u \DP{\bar{v}}{x} + (\bar{v} - \frac{3}{2}T 
        \Omega_K x)\DP{\bar{v}}{y} + w\DP{\bar{v}}{z} 
      = - \frac{1}{\rho} \DP{P}{y}  
	- (2C - \frac{3}{2} T) \Omega_{K} u -\DP{\Psi}{y},
	\label{eqn:bta4}
\end{equation}
\begin{equation}
	 \DP{w}{t} + u \DP{w}{x} + 
 	(\bar{v} - \frac{3}{2} T\Omega_K x)  \DP{w}{y} +w \DP{w}{z} 
      = - \frac{1}{\rho} \DP{P}{z} -  \Omega_K ^2 z -\DP{\Psi}{z}, 
	\label{eqn:bta5}
\end{equation}
 \begin{equation}
	   \DPt{\Psi}{x} + \DPt{\Psi}{y} + \DPt{\Psi}{z} 
       =  4 \pi G \rho.
	 \label{eqn:bta6}
 \end{equation}

In order to investigate the linear stability, 
we assume that an unperturbed state is steady and uniform 
in $x$ and $y$ directions:
\begin{equation}
 \DP{}{t}= \DP{}{x}= \DP{}{y}= 0,
	\label{eqn:ntp.1}
\end{equation}
except for $\partial P_0/\partial x =$constant$ \neq  0$. 
We also assume that the unperturbed velocity is parallel 
to the $y$-axis (i.e. the azimuthal direction)
\begin{equation}
  u_0 = w_0 = 0, 
	\label{eqn:ntp.2}
\end{equation}
and $\rho_g=$constant throughout the region of consideration. 
The unperturbed azimuthal velocity is determined from 
%MS% equations 
equation %MS%
(\ref{eqn:unv}), which is rewritten 
\begin{equation} 
   \bar{v}_0=-\rho_g\eta v_K/\rho_0, 
   \label{eqn:v0bar} 
\end{equation} 
where $\eta$ is assumed to be constant in our local Cartesian coordinates. 

From equations (\ref{eqn:bta3}), (\ref{eqn:ntp.1}) and (\ref{eqn:ntp.2}), 
we have 
\begin{equation}
 \frac{1}{\rho_0}\DP{P_0}{x} = 2  C \Omega_K {\bar{v}_0}.
	\label{eqn:ntp.3}
\end{equation}
Integrating equation (\ref{eqn:bta6}) with equation (\ref{eqn:ntp.1})
we have 
\begin{equation} 
	\DP{\Psi_0}{z} = 4 \pi G \sigma_0,
	\label{eqn:ntp.4}
\end{equation}
where 
\begin{equation} 
	\sigma_0  =  \int^{z}_{0} \rho_0 dz.
	\label{eqn:ntp.5}	
\end{equation}
From equations (\ref{eqn:bta5}), (\ref{eqn:ntp.2}) and (\ref{eqn:ntp.4}),
 we obtain
\begin{equation} 
\frac{1}{\rho_0} \DP{P_0}{z} = - \Omega_K^2 z - 4\pi G \sigma_0.
	\label{eqn:ntp.6}
\end{equation}

Linearizeing equations (\ref{eqn:bta1}) to (\ref{eqn:bta6})
and using equations (\ref{eqn:ntp.3}), (\ref{eqn:ntp.4}), 
(\ref{eqn:ntp.6}), we have
\begin{equation}
	 \DP{ u_1}{x} + \DP{v_1}{y} + \DP{w_1}{z} =0,
	\label{eqn:tpe.1}
\end{equation}
\begin{equation}
	\DP{ \rho_1}{t} + (\bar{v}_0-\frac{3}{2}T \Omega_Kx)\DP{\rho_1}{y}
+\DD{\rho_0} {z}w_1=0,
		\label{eqn:tpe.2}
\end{equation}
\begin{equation}
	  \DP{ u_1}{t}+(\bar{v}_0-\frac{3}{2}T \Omega_K x)\DP{u_1}{y}
      = - \frac{1 }{\rho_0} \DP{P_1}{x} + 2 C \Omega_K \frac{\bar{v}
         _0}{\rho_0}\rho_1 + 2C  \Omega_{K}v_1 - \DP{\Psi_1}{x},
		\label{eqn:tpe.3}
\end{equation}
\begin{equation}
       \DP{ v_1}{t}+(\bar{v}_0-\frac{3}{2}T \Omega_K x)\DP{v_1}{y}
	+\DD{\bar{v}_0}{z}w_1 
      = - \frac{1}{\rho_0} \DP{P_1}{y} - (2C-\frac{3}{2}T) \Omega_{K}u_1 
	- \DP{\Psi_1}{y}, 
		\label{eqn:tpe.4}
\end{equation}
\begin{equation}
 \DP{ w_1}{t}+(\bar{v}_0-\frac{3}{2}T \Omega_K x)\DP{w_1}{y}
  = - \frac{ 1}{\rho_0} \DP{P_1}{z}
  + \frac{1}{\rho_0^2}\DP{P_0}{z} \rho_1 
  - \DP{\Psi_1}{z}, 
		\label{eqn:tpe.5}
\end{equation}
\begin{equation}
	   \DPt{\Psi_1}{x} + \DPt{\Psi_1}{y} + \DPt{\Psi_1}{z} 
       =  4 \pi G \rho_1, 
		\label{eqn:tpe.6}
\end{equation}
where a perturbed quantity is denoted by subscript 1.

\section{The Effect of the Coriolis Force}
\subsection{Eigenvalue Problem}
First, we consider the case without the self-gravity and tidal force, namely, 
$G=0$, $\Psi_1=0$ 
and $T=0$ in order to investigate the effect of the Coriolis force.
Then we can assume that a perturbed quantity $f_1$ has the form as 
\begin{equation}
   f_1(x,y,z,t) = \hat{f}_1(z) \exp[i(k_x x + k_y  y- \omega t)],
	\label{eqn:pef}
\end{equation}
where 
$\omega$ is the complex angular frequency of perturbed quantities,
$k_x$ is the radial wave number, and $k_y$ is the azimuthal wave number.
Note that $\hat{f}_1(z)$ is a function of $z$. 
By assuming that all the perturbed quantities have the form as equation (\ref{eqn:pef}), and the self-gravity and the tidal force are neglected, equations 
(\ref{eqn:v0bar}) and (\ref{eqn:tpe.1})--(\ref{eqn:tpe.5}) are 
rewritten using $\Psi_1=0$, $T=0$ 
and equations (\ref{eqn:ntp.6}) with $G=0$ and (\ref{eqn:pef})
 (we omit $\hat{}$  in the following equations)
\begin{equation} 
	 v_0=-\rho_g\eta v_K/\rho_0, 
	\label{eqn:v0} 
\end{equation} 
\begin{equation}
	 i k_x u_1 + i k_y v_1  + \DD{w_1}{z} =0,
	\label{eqn:cpe.1}
\end{equation}
\begin{equation}
	 - i \bar{\omega} \rho_1+\DD{\rho_0} {z}w_1=0,
	\label{eqn:cpe.2}
\end{equation}
\begin{equation}
      -	i \bar{\omega} u_1 
      = - i k_x\frac{1 }{\rho_0} P_1 + 2 C \Omega_K \frac{v_0}{\rho_0}\rho_1
	+ 2 C\Omega_{K}v_1,
	\label{eqn:cpe.3}
\end{equation}
\begin{equation}
 -i \bar{\omega} v_1 
      = - i k_y \frac{1}{\rho_0} P_1 - \DD{v_0}{z} w_1 
 - 2 C \Omega_{K}u_1,
	\label{eqn:cpe.4}
\end{equation}
\begin{equation}
 -	i \bar{\omega} w_1 
  = - \frac{ 1}{\rho_0} \DD{P_1}{z}
  - \frac{\Omega_K ^2  z}{\rho_0} \rho_1,
	\label{eqn:cpe.5}
\end{equation}
where
\begin{equation}
\bar{\omega} = \omega - k_y v_0(z).
\end{equation}
From Eq. (\ref{eqn:cpe.2}), we have
\begin{equation}
\rho_1 = \frac{1}{i \bar{\omega}} \DD{\rho_0}{z} w_1.
 	\label{eqn:crho}
\end{equation}
From equation (\ref{eqn:v0}), we have
\begin{equation}
     \DD{v_0}{z} = -\frac{v_0}{\rho_0} \DD{\rho_0}{z},
 	\label{eqn:dv_0}
\end{equation}
and 
\begin{equation}
     \DDt{v_0}{z} = v_0\left[ 2\left(\frac{1}{\rho_0}\DD{\rho_0}{z}\right)^2 
     - \frac{1}{\rho_0}\DDt{\rho_0}{z}\right]. 
	\label{eqn:d2v_0}	
\end{equation}
From equations (\ref{eqn:cpe.3}), (\ref{eqn:cpe.4}),  (\ref{eqn:crho})
and (\ref{eqn:dv_0}), we have
\begin{equation}
u_1 = \frac{k_x \bar{\omega} + 2 i C k_y \Omega_{K}}
{\bar{\omega}^2 - 4 C^2 \Omega_{K}^2}\frac{P_1}{\rho_0},
	\label{eqn:u1}
\end{equation}
and
\begin{equation}
v_1 = \left[(k_y \bar{\omega} - 2 i  C k_x \Omega_{K})\frac{P_1}{\rho_0}
 + \left(- i \bar{\omega}\DD{v_0}{z} 
+\frac{ 4 C^2\Omega_{K}^2}{i \bar{\omega}} \frac{v_0}{\rho_0}
\DD{\rho_0}{z} \right) w_1 \right] 
\left( \bar{\omega}^2 - 4 C^2\Omega_{K}^2 \right)^{-1}.
	\label{eqn:v1}
\end{equation}
Substituting equations (\ref{eqn:u1}) and (\ref{eqn:v1})  
into equation (\ref{eqn:cpe.1}), we have
\begin{equation}
P_1= \frac{i \rho_0}{(k_x^2+k_y^2)^2 \bar{\omega}} %MS%
 \left[ (\bar{\omega}^2 - 4 C^2 \Omega_{K}^2) \DD{w_1}{z}
 + k_y \left( \bar{\omega}\DD{v_0}{z}
 + \frac{4 C^2 \Omega_{K}^2}{\bar{\omega}}\frac{v_0}
{\rho_0}\DD{\rho_0}{z} \right) w_1 \right].
\label{eqn:p1}
\end{equation}
Substituting equation (\ref{eqn:crho}) and (\ref{eqn:p1}) 
into ({\ref{eqn:cpe.5}), and using equations (\ref{eqn:dv_0}) and
(\ref{eqn:d2v_0}), we get
\begin{equation}
	\DDt{w_1}{z} + E \DD{w_1}{z} + F w_1= 0,
	\label{eqn:cb1}
\end{equation}		
where
\begin{equation}
E = \frac{1}{\rho_0} \DD{\rho_0}{z}  
      \left[1 + 
      \frac{8 C^2 \Omega ^2 k_y v_0}
      {\left(\bar{\omega}^2-4C^2\Omega^2 \right)\bar{\omega}}\right], %MS%
	\label{eqn:cb2}
\end{equation}	
\begin{eqnarray}
 F&=& \frac{k_y v_0}{\bar{\omega}} \left[\left(\frac{1}{\rho_0}\DD{\rho_0}{z}\right)^2
      -\frac{1}{\rho_0}\DDt{\rho_0}{z} \right] \nonumber\\
   && -\frac{1}{\bar{\omega}^2 - 4 C^2\Omega_{K}^2}
	\left[\left(k_x^2+k_y^2\right)^2 \left( \bar{\omega}^2 + \frac{1}{\rho_0}\DD{\rho_0}{z}\Omega_K^2 z\right)
	+8\left(\frac{C\Omega_K k_y v_0}{\bar{\omega} \rho_0}\DD{\rho_0}{z}\right)^2 \right].
	\label{eqn:cb3}
\end{eqnarray}
Thus, perturbation equations (\ref{eqn:tpe.1})--(\ref{eqn:tpe.5}) are 
reduced to a differential equation (\ref{eqn:cb1}) for $w_1$ 

Only odd solutions for $w_1$ are considered because 
even ones are always stable according to our calculations.
Thus one of the boundary conditions is given by 
\begin{equation}
 w_1 = 0  \; \mbox{at}  \;  z= 0.
 \label{eqn:bcin} 
\end{equation}
Outside the dust layer, from equation
 (\ref{eqn:cb1}) to (\ref{eqn:cb3}), we have
\begin{equation}
\DDt{w_1}{z} -K^2 w_1=0,
\end{equation}
where
\begin{equation}
K^2 = \frac{
%MS% k^2
\left(k_x^2+k_y^2\right) %MS%
\bar{\omega}^2}{\bar{\omega}^2 - 4C^2 \Omega_K^2}.
\end{equation}
We select a root $K$ whose real part is positive.
Then the outer boundary condition, {\it i.e.}, 
$w_1 \rightarrow 0 \; \mbox{for}  \;   z \rightarrow \infty,$
is satisfied by the solution,
\begin{equation}
	w_1 \propto \exp(-K z).
	\label{eqn:bwbw}
\end{equation}
From equation (\ref{eqn:bwbw}), we have
\begin{equation}
	\DD{w_1}{z} + K w_1=0 \; \mbox{for}  \; z>z_d .
	\label{eqn:bcb}
\end{equation}
From equations (\ref{eqn:p1}) and (\ref{eqn:bcb}), we get
\begin{equation}
P_1 = \frac{i\rho_g(\bar{\omega}^2 -4 C \Omega_K ^2)}{
\left(k_x^2+k_y^2\right) 
\bar{\omega}}
\DD{w_1}{z} =
-\frac{i\rho_g(\bar{\omega}^2 -4C \Omega_K^2)K}{
\left(k_x^2+k_y^2\right) 
\bar{\omega}} w_1. 
\label{eqn:bp1}
\end{equation}
At the boundary between the dust and the gas layers, 
$P_1$ and $w_1$ must be continuous.
Thus equations 
(\ref{eqn:p1}) and (\ref{eqn:bp1}) read
by using equation (\ref{eqn:dv_0}) 
\begin{eqnarray}
\DD{w_1}{z}+ \left(K-\frac{k_yv_0}{\bar{\omega}\rho_0} 
\DD{\rho_0}{z}\right)w_1 =0 
\; \mbox{at}  \;  z= z_d.
 \label{eqn:bcout}
\end{eqnarray}
We obtain eigenvalues $\omega$ by solving 
differential equation (\ref{eqn:cb1}) to satisfy the boundary 
conditions (\ref{eqn:bcin}) and (\ref{eqn:bcout}). 
The growth rate of the shear instability is $\omega_I = \Im(\omega)$, where $\Im$ denotes the imaginary part.

In this work, we employ CRNDDD unlike \citet{ishitsu02} which used HDDD. 
However, CRNDDD is not an explicit function of $z$ 
(see eq. [\ref{eqn:Richard}]).
Thus, we integrate the differential equation by using the dust surface density $\sigma_0$ as the independent variable.
Then we have to calculate the density $\rho_0$ and the height $z$ 
as functions of the surface density $\sigma_0$.
From equations (\ref{eqn:u}) and 
(\ref{eqn:ntp.5}) with $\rho_0=\rho_d+\rho_g$, 
we have
\begin{equation}
   \sigma_0  = \rho_g \int^{u}_{u_m} \frac{1}{u-q} \frac{\partial z}{  \partial u} du. 
	\label{eqn:sigma0}	
\end{equation}
Differentiating equation (\ref{eqn:Richard}) with respect to $u$ and substituting 
in equation ({\ref{eqn:sigma0}) and integrating, 
we get 
\begin{equation}
\sigma_0 = \sqrt{J}\eta r \rho_g 
\ln 
\left[ \left( u/u_m \right) 
+ \sqrt{(u/u_m)^2 -1} \right]. 
\label{eqn:sigmau}
\end{equation}
This equation is solved for $u$ as 
\begin{equation} 
u=u_m \cosh \left(\sigma_0/\sqrt{J} \eta r \rho_g \right). 
\label{eqn:uapa}
\end{equation}
Next, we obtain $\rho_0$ using 
equation (\ref{eqn:u}), 
\begin{equation}
	\rho_0 = \rho_g [u_m \cosh (\sigma_0 / \sqrt{J} \eta r \rho_g ) - q]^{-1}
	\label{eqn:rhos} 
\end{equation}
We can get $z$ from equations (\ref{eqn:Richard}) using equation (\ref{eqn:uapa}), 
and $v_0$ from equation (\ref {eqn:v0}) using equation  (\ref{eqn:rhos}).
Differentiating equation (\ref{eqn:rhos}) by $z$ and using equation (\ref{eqn:ntp.5}), we get
\begin{equation}
\DD{\rho_0}{z} = -\rho_0^3 \frac{u_m}{\sqrt{J} \eta r \rho_g^2} 
\sinh(\sigma_0 / \sqrt{J} \eta r \rho_g ),
\end{equation} 
and
\begin{equation}
\DDt{\rho_0}{z} = \frac{3}{\rho_0} \left(\DD{\rho_0}{z} \right)^2 
- \frac{\rho_0^4 u_m}{J \eta^2 r^2 \rho_g^3} \cosh(\sigma_0 / \sqrt{J} \eta r \rho_g).
\end{equation}
Thus the 
differential equation (\ref{eqn:cb1}) and the boundary condition (\ref{eqn:bcout}) are rewritten
\begin{equation}
	\DDt{w_1}{\sigma_0} + \frac{1}{\rho_0^2} \left(\DD{\rho_0}{z} + \rho_0
    E \right) 
	\DD{w_1}{\sigma_0} + \frac{F}{\rho_0^2} w_1 =0, 
	\label{eqn:difeq}
\end{equation}
and 
\begin{eqnarray}
\DD{w_1}{\sigma_0}+ \frac{1}{\rho_0}
\left(K-\frac{k_yv_0}{\bar{\omega}\rho_0} \DD{\rho_0}{z}\right)w_1 =0 
\; \mbox{at}  \;  z= z_d.
\label{eqn:bund}
\end{eqnarray}
If there are an eigenvalue $\omega$ and an eigenfunction $w_1$ of 
the differential equation (\ref{eqn:difeq}) under the boundary conditions (\ref{eqn:bcin}) and (\ref{eqn:bund}), 
then their complex conjugates $\omega^*$ and $w_1^*$ also satisfy these equations as seen by taking complex conjugates of these equations. 
Thus, if there is a growing mode with the growth rate $\omega_I$, then there is also a decaying mode with $-\omega_I$, and {\it vice versa}. 
Replacing $k_y$ by $-k_y$ and $\omega$ by $-\omega$ in equations (\ref{eqn:difeq}) and (\ref{eqn:bund}), we get the identical equations. 
Thus, we have a same growth rate of growing modes for opposite signs of $k_y$ with a same absolute value. 
Further it is easily seen that eigenvalues are same for opposite signs of $k_x$ with a same absolute value. 
Consequently, the growth rate of an unstable mode is an even function of $k_x$ and $k_y$.

\subsection{Results}
The Richardson number $J$ and the dust density at the midplane $\rho_d(0)$ were taken as independent parameters in \citet{sekiya00} on determining CRNDDD. 
However, the Richardson number $J$ is determined if 
the dust density at the midplane $\rho_d(0)$ and the dust surface density factor 
$f_d$ are given for CRNDDD. 
Figure \ref{fig1} draws the Richardson number as functions 
of the dust density at the midplane $\rho_d(0)$ with constant dust surface densities, which are calculated using equation (\ref{eqn:sigmad}).

Figure \ref{fig2} shows the growth rate $\omega_I$ as a function of 
the absolute values of radial and azimuthal wave numbers $k_x$
 and $k_y$ in the case where $\rho_d(0) /\rho_g= 1.0$, 
and $f_d=1.0$ at $R=$1AU.
Hereafter, non-dimensional wave numbers, $k_x\eta R$ and $k_y\eta R$, are used instead of the dimensional ones, $k_x$ and $k_y$, where $\eta R$=constant in our local Cartesian coordinate system. 
This figure shows that there exists a maximum radial wave number for the shear instability $k_{xc} (>0)$ for each value of $k_y$; if $|k_x|>k_{xc}$, the growth rate $\omega_I=0$. 
Figure \ref{fig3} shows the growth rate $\omega_I$ as a function of 
$|k_x|\eta R$ with keeping $|k_y|\eta R=$10.5 
in the case where $\rho_d(0) /\rho_g= 1.0$, and $f_d=1.0$ 
(same as Fig. \ref{fig2}). 
We have $k_{xc}\eta R=$190 from this figure. 
Figure \ref{fig4} shows a graph similar to Figure \ref{fig3}, 
but for $\rho_d(0) /\rho_g= 40$ and $|k_y| \eta R=$180; 
in this case, we have $k_{xc}\eta R=$ 340. 
These figures will help us to understand the stabilization and amplification mechanisms of the shear instability by the tidal force stated in \S 5. 

Figure \ref{fig5} shows the growth rate of the mode with the most unstable wave number (hereafter referred to as ``the peak growth rate") as a function of the dust/gas density ratio on the midplane. 
In the case without the Coriolis force, the peak growth rate is of the order of the Kepler angular velocity $\Omega_K$ over the entire dust density and increases monotonically for density smaller than $\rho_d(0)/\rho_g=50$ except for $2<\rho_d(0)/\rho_g<10$. 
On the other hand, the peak growth rate decreases for $\rho_d(0)/\rho_g>50$. 
In the case with the Coriolis force, the peak growth rate is smaller than that without the Coriolis force. 
Moreover, for $\rho_d(0)/\rho_g>100$, the shear instability is stabilized! 

This is explained as follows. 
For a system without the Coriolis force, 
the energy of instability 
is supplied from the unperturbed flow at the co-rotation sheet, where the angular phase velocity of the perturbed quantities is coincident with the unperturbed azimuthal velocity (see \citealt{sekiya00} for detail).  
However, when the Coriolis force is included, other resonances like the Lindblad resonances appear. 
In the system with the Coriolis force, perturbed energy is supplied 
from the latter ``resonances" by the baroclininic instability instead of the co-rotation resonance. 
Figure \ref{fig6} shows that the real and imaginary parts of the eigenfunction $\Re(w_1)$ and $\Im(w_1)$, respectively, in the case where 
$k_y \eta R=32.4$, $k_x =0$,  $\rho_{d}(0)/\rho_g=12$, $C=1$ and $f_d=1$.
The positions of the co-rotation resonance, and the upper and the lower 
resonances are presented in Figure \ref{fig6}.  
Figure \ref{fig7} also shows the eigenfunctions in the case with a higher density where $k_y \eta R=180$, $k_x =0$,  $\rho_{d}(0)/\rho_g=40$, $C=1$ and $f_d=1$. 
Compared to Figure \ref{fig6}, the intervals between the resonances become narrower as the dust density at midplane increases.
The amplitude of an eigenfunction largely changes around the co-rotation, the upper and the lower resonances. 
It becomes more difficult to connect the eigenfunction as $\rho_{d}(0)/ \rho_g$ increases because the intervals between the resonances decrease. 
Thus, no eigenvalues exist for $\rho_d(0)/\rho_g \geq 100$, namely the shear instability is stabilized.

\section{The Effects of the Tidal Force and the Self-Gravity} 
In this section the linear perturbation equations are solved taking both the tidal force and the self-gravity into account as well as the Coriolis force. 
The tidal force plays a crucial role for the stabilization of 
the shear instability.

\subsection{Formulation} 
 The normal mode analysis ({\it i.e.} the Fourier transform) with respect to
$x$ cannot be done in the coordinate system written in the previous section 
because the coefficients of equations depend on $x$, if $T\ne 0$. 
Thus we transform the coordinate $y$ into the coordinate $y'$ shearing 
with the local Kepler velocity $-\frac{3}{2} T \Omega_K x$ 
(see Eq. [\ref{eqn:keplerv}])
in order that the normal mode analysis for $x$ can be done, 
\begin{equation}
   y'= y + \frac{3}{2}T \Omega_K x t.
\end{equation}
We assume that perturbed quantities are written
\begin{equation}
   f_1(x,y,z,t) = \hat{f}_1(z,t) \exp[i (k_y  y'+ k_x x)].
	\label{eqn:exp}
\end{equation}
Substituting this expression into equations (\ref{eqn:tpe.1})--(\ref{eqn:tpe.6}), 
and abbreviating $\hat{f}_1(z,t)$ as $f_1$,
the perturbation equations are rewritten
\begin{equation}
	 ik'_xu_1 + i k_y v_1 + \DP{w_1}{z} =0,
	\label{eqn:pt1}
\end{equation}
\begin{equation}
	\DP{ \rho_1}{t} +  i k_y \bar{v}_0 \rho_1
+\DD{\rho_0} {z}w_1=0,
	\label{eqn:pt2}
\end{equation}
\begin{equation}
	  \DP{ u_1}{t}+ i k_y \bar{v}_0  u_1
      = - i k'_x \frac{1}{\rho_0} P_1 + 2C \Omega_K 
       \frac{\bar{v}_0}{\rho_0}\rho_1 + 2 C\Omega_{K}v_1 - i k'_x \Psi_1,
	\label{eqn:pt3}
\end{equation}
\begin{equation}
       \DP{ v_1}{t}+ i k_y \bar{v}_0  v_1
      = - i k_y \frac{1}{\rho_0} P_1  - \DD{\bar{v}_0}{z} w_1
      - \left(2C - \frac{3}{2}T\right) \Omega_{K} u_1 - i k_y \Psi_1,
	\label{eqn:pt4}
\end{equation}
\begin{equation}
 \DP{ w_1}{t}+ i k_y \bar{v}_0  w_1
  = - \frac{ 1}{\rho_0} \DP{P_1}{z}
+ \frac{1}{\rho_0^2} \DP{P_0}{z}
\rho_1 -  \DP{\Psi_1}{z},
	\label{eqn:pt5}
\end{equation}
\begin{equation}
 	\DPt{\Psi_1}{z}   - (k'^{2}_x + k^{2}_y)  \Psi_1
=  4 \pi G \rho_1,
	\label{eqn:pt6}
\end{equation}
where,
\begin{equation}
	 k'_x = k_x +  \frac{3}{2}T k_y \Omega_K t.
	\label{eqn:pt7}
\end{equation}
The coefficients of these equations depend on $t$.
Thus, the normal mode analysis (i.e. the Fourier transform with respect to $t$) cannot be performed in the case with the tidal force. 
Additionally, the self-gravity brings out an extra free parameter when obtaining an eigenvalue, so that the calculation becomes difficult. 
Thus, we numerically integrate equations (\ref{eqn:pt1}) to 
(\ref{eqn:pt7}) following the method of \citet{ishitsu03} (see Appendix A).

It must be noted that some unperturbed quantities such as $\rho_0$ and $v_0$ are not differentiable at $z_d$ for CRNDDD 
(see Figs. 1 to 3 in \citealt{sekiya00}).  
Accordingly our equations become singular there. 
In order to avoid it, we soften density distribution around $z=z_d$ as follows.
We choose a point $z_c$ which satisfies $0\ll z_c<z_d$. 
For $z<z_c$, we use CRNDDD; on the other hand, for $z>z_c$, 
dust density is assumed 
\begin{equation}
	\rho_d(z) = A \exp[ -B(z-z_c)],
\end{equation}
where constants $A$ and $B$ are determined so that the dust density is continuous and differentiable at $z=z_c$. 
In present paper, we put $z_c=0.95z_d$ in all the calculations.
Here we adopted this value of $z_c$ so that the value dose not affect the results and our calculations do well. 
Actually we confirmed that our code worked well when $z_c =0.95$. 
Then the Richardson number for $z>z_c$ is always larger than the constant Richardson number for $z<z_c$.

Boundary conditions for the $z$-direction are given as follows.
The mirror symmetry with respect to the midplane is assumed: 
\begin{equation} 
 w_1 = 0  \; \mbox{ at }  \;  z= 0, 
	\label{eqn:bwm} 
\end{equation} 
\begin{equation}
 \DP{P_1}{z} = 0  \; \mbox{ at }  \;  z= 0, 
	\label{eqn:bpm} 
\end{equation} 
and
\begin{equation}
 \DP{\Psi_1}{z} = 0  \; \mbox{ at }  \;  z= 0. 
	\label{eqn:bpsim}
\end{equation}
The continuity of the pressure at $z=z_d$, i.e. 
the boundary between the dust and the gas layers
is applied in the case without the tidal force (see \S 4). 
However, it is difficult to apply this condition to the case with the 
tidal force.
Thus, we solve the perturbation equations numerically within
 region $[0,z_0]$, where the solid-wall condition is applied  
at $z_0$ for simplicity: 
\begin{equation} 
 w_1 = 0  \; \mbox{ at }  z= z_0. 
	\label{eqn:bwz0} 
\end{equation} 
The value of $z_0$ is chosen to be large enough 
in order for an eigenfunction to decay sufficiently at a boundary $z_0$.
We used $z_0 = 2 z_d$ in our calculations, and confirmed 
that numerical solutions for $T=0$ with this condition 
 agree well with  the results of the normal mode analysis in \S 4.
From equations 
(\ref{eqn:pt2}), (\ref{eqn:bwm}) and (\ref{eqn:bwz0}), 
we have
\begin{equation}
 \DP{\rho_1}{t} + i k_y \bar{v}_0 \rho_1= 0 
 \; \mbox{ at }  \;  z= 0 \mbox{ and } z_0.
\label{eqn:brho}
\end{equation}
If we give an initial condition with  $\rho_1=0$ at $z=0 \mbox{ and } z_0$,
equation (\ref{eqn:brho}) implies that $\rho_1=0$ for $ t \geq 0 $.
Hereafter we take only such initial conditions for simplicity.
 Accordingly boundary conditions for $\rho_1$ are given by
\begin{equation}
\rho_1= 0  \; \mbox{ at }  \;  z= 0 \mbox{ and } z_0.
	\label{eqn:bt3}
\end{equation}
By assuming that $\rho_1=0$ for $z> z_0$, equation (\ref{eqn:pt6}) reads 
\begin{equation}
 	\DPt{\Psi_1}{z}   - k'^{2}  \Psi_1 =0,
	\label{eqn:bphi1}
\end{equation}
where 
\begin{equation}
k' =  \sqrt{k'^{2}_x + k^{2}_y}. 
\end{equation}
The solution of (\ref{eqn:bphi1}), which satisfies 
$\Psi_1 \rightarrow 0 $ for $z \rightarrow 0$, is given by 
\begin{equation}
 	\Psi_1 \propto  \exp( -k' z).
	\label{eqn:bphi2}
\end{equation}
From equation (\ref{eqn:bphi2}), we have 
the boundary condition for $\Psi_1$ at $z=z_0$: 
\begin{equation}
  \DP{\Psi_1}{z} = - k' \Psi_1  \; \mbox{ at }  \;  z= z_0. 
      \label{eqn:bt4}
\end{equation}
From (\ref{eqn:pt5}), (\ref{eqn:bwz0}), (\ref{eqn:bt3}) and  (\ref{eqn:bt4}), 
the boundary condition for $P_1$ at $z=z_0$ is given by
\begin{equation}
 \DP{P_1}{z} = k' \rho_0 \Psi_1  \; \mbox{ at }  \;  z= z_0.
	\label{eqn:bt6}
\end{equation}

\subsection{The Effect of the Self-Gravity}
Figure \ref{fig8} shows the results neglecting both the Coriolis force and the tidal force in order to see the effects of the self-gravity of the dust layer.
The self-gravity of the dust layer does not affect the shear instability as much as the Coriolis force and the tidal force.
The self-gravity decreases 
the growth rate of the shear instability somehow.
It is natural to expect that the self-gravity has little effects 
on the shear instability by noting that the growth rate of the gravitational
 instability is  highest for a much smaller radial wave number $k_x$. 
Thus, the self-gravity does not change the results of the shear instability. 

\subsection{The Effect of the Tidal Force} 
Here, we mainly investigate the stability of shear flow 
for $\rho_d(0)/\rho_g<100$
for which the Coriolis force cannot stabilize the shear instability (see \S 4).
For $\rho_d(0)/\rho_g = 40$, Figure \ref{fig9} shows the time evolution of the 
radial, azimuthal and vertical components of the spatially averaged kinetic energy of the perturbed flow; the values of the parameters in Figure \ref{fig9} are same as Figure \ref{fig4}, 
except that $T=1$ in Figure \ref{fig9}, while $T=0$ in Figure \ref{fig4}. 
These components vibrate, and the growth rate of the shear instability is effectively zero.
Thus, the tidal force stabilizes the instability.
The calculations for several other parameters have been performed, and the tidal force always stabilizes the instability.

This stabilization is explained by the following mechanism (see also \citealt{ishitsu03}). If the 
tidal force does not work ($T=0$), we can obtain the growth rate of the shear instability with the method written in section 4. 
The growth rate depends on the radial and the azimuthal wave numbers $k_x$ and $k_y$. 
The radial wave number plays a crucial role. 
There exists a critical radial wave number $k_{xc}$ ($>0$); if $|k_x|>k_{xc}$, then 
the growth rate is zero (see Figs. \ref{fig2} to \ref{fig4}). 
If the tidal force is included, that is, the radial shearing works, the radial wave number changes with time due to the Keplerian shear: 
\begin{equation}
    k'_x = k_x+\frac{3}{2}T k_y t \Omega_K.
	\label{kxt}
\end{equation}
If $|k_x|<k_{xc}$, the perturbed quantities grow at first. 
As time passes, the value of $|k'_x|$ changes, and eventually $|k'_x|$ 
exceeds the critical wave number $k_{xc}$, then 
the growth rate becomes zero, that is, the shear instability is stabilized.
From equation (\ref{kxt}), we evaluate the stabilization time $t_s$ 
by 
\begin{equation}
    t_s = \left[\mbox{sgn}(k_y)k_{xc}-k_x \right]/\left(\frac{3}{2}T k_y \Omega_K \right). 
\end{equation}
In the case of Figure \ref{fig4}, $k_{xc}\eta R  =340$ and
we have $t_s= 1.3 \Omega_K^{-1}$ for $k_x=0$. 
As seen in Figure \ref{fig9}, the perturbation energy does not grow for $t>t_s$. 
 
Additionally, 
the change of the radial wave number with time has not only the 
stabilizing effect but also amplification mechanism in some occasions. 
Figures \ref{fig10} and \ref{fig11} show
time evolution of perturbation energy and the growth rate, respectively, 
in the case where $k_x\eta R=-190$ and 
$k_y\eta R=10.5$ (same values of the parameters as Figure \ref{fig3}, 
except that $T=1$ in Figures \ref{fig10} and \ref{fig11}, 
while $T=0$ in Figure \ref{fig3}). 
The growth rate 
oscillates, but the averaged value for each oscillation 
increases at first ($t\Omega_K<$12, i.e. $k'_x<$0). 
After that time, however, it begins to decrease 
(12$< t\Omega_K<$24, i.e. 0$<k'_x\eta R<$190), and finally it stops growing 
(24$<t\Omega_K$, i.e. 190$<k'_x\eta R$). 
These changes of the growth rate of the energy can be interpreted by considering the change of the radial wave number in Figure \ref{fig3}. 
If the value of $k_x$ increases from the initial value 
$k_x\eta R=-190$ with keeping $k_y\eta R=10.5$, 
the growth rate first increases, and reaches a peak value at $k_x=0$; 
after that, the growth rate decreases, 
and finally, the mode is stabilized for $k_x \eta R >k_{xc}\eta R=$190. 
The increase of the growth rate during 0$<t\Omega_K<$12 (i.e. $k'_x<$0) in Figure \ref{fig9} resembles the `swing amplification' \citep{goldreich65} in the sense that the Keplerian shear causes the growth. 
However, our equations are different from \citet{goldreich65} in the points that 
the fluid is incompressible and the self-gravity plays no important roles in our case. 
Although the growth rate increases for adequate ranges of $k'_x$, 
the stabilization effect would consequently overcome the amplification. In fact, we confirmed by a number of numerical simulations that the amplification has little effect on our results.

In the case without the tidal force, the growth rate is of the order of the Kepler angular velocity. 
The instability with such a small growth rate is stabilized by the tidal force \citep{ishitsu03}. 
We have shown in this subsection that CRNDDD is stable against the shear instability for $\rho_d(0)/\rho_g<100$.  
Moreover, CRNDDD is stabilized due to the Coriolis force for $\rho_d(0)/\rho_g>100$ as shown in \S 4.
Thus, CRNDDD is stable for all range of the dust density at the midplane.

\section{Discussion and Conclusions}

Recently, \citet{ishitsu03} presented that the tidal force has the effect to stabilize the shear instability.
They used the constant dust distribution with sinusoidal transitional 
zones as an unperturbed state, which corresponds to an initial distribution before the onset of the shear instability.
They showed that the shear instability develops before the onset of the gravitational instability because the growth rate of the shear instability becomes so large that the Coriolis and the tidal forces can not stabilize it. 
Thus, the dust layer would become turbulent.  
The dust density distribution changes due to the turbulent diffusion and the Richardson number is expected to approach a constant value.  
In this paper, we used the constant Richardson number dust density distribution as an unperturbed state and investigated its linear stability.

First, we showed that the Coriolis force stabilizes the shear instability if the dust density on the midplane is larger than a hundred times the gas density. 
Next, we found that the coupling effect of the tidal and the Coriolis forces stabilizes the shear instability also for the midplane dust density less than a hundred times the gas density. 
Thus, the constant Richardson number dust density distribution is always stable against the shear instability.
Dust settling proceeds, and eventually the dust density on the midplane exceeds the critical density of the gravitational instability. 
It is possible that the gravitational instability of the dust layer occurs even for the dust to gas surface density ratio is that for the solar composition. 
So far, it has been considered that the gravitational instability requires the special mechanisms of the dust concentration or the gas dissipation in order to overcome the shear instability \citep{sekiya98, youdin02, youdin03}. 
The stabilizing 
%MS% effect 
effects %MS%
of the tidal and 
the %MS%
Coriolis forces elucidated in this paper relaxes this restriction. 
And the gravitational instability may be a common process of the planetesimal formation.

In this paper, we assumed that the constant dust density distribution as an unperturbed state, and showed that it is stable.  
In the real evolution of the dust layer, the dust settling and the turbulent mixing may 
%MS% cuase 
cause %MS%
the dust distribution somewhat different from that with the constant Richardson number.  
It is also unclear yet whether the planetesimals really form after the dust density exceeds the critical density of the gravitational instability. 
In order to elucidate the actual evolution of the dust layer, we plan to perform nonlinear, dust-gas multifluid numerical simulation in future.

\acknowledgments
We thank Dr. Seiichiro Watanabe for valuable comments. The calculations in this work were partly performed with computers at Astronomical Data Analysis Center, National Astronomical Observatory of Japan. 



\appendix
%MS% \section{Detail of Numerical Method}
\section{Numerical Method}
We adopt the MAC method, in which the pressure is determined by the condition that the equation of continuity is satisfied in the next step, and other variables $u_1, v_1, w_1$ and $\rho_1$ are calculated using this value of the pressure. 

We define the divergence of the perturbed velocity by
\begin{equation}
	D \equiv ik'_x u_1 + i k_y v_1 + \DP{w_1}{z}.
	\label{eqn:div}
\end{equation}
Multiplying equations (\ref{eqn:pt3}) by $ i k'_x $ and equation
(\ref{eqn:pt4}) by $i k_y$, and taking partial derivative of equation (\ref{eqn:pt5}) with respect to $z$, and adding, we have
\begin{eqnarray}
	\DP{D}{t} &=& 
- i k_y \bar{v}_0 D - 2 i k_y \DD{\bar{v}_0}{z} w_1  
-\frac{1}{\rho_0} \left( -k'^2_x -k^2_y  
+  \DPt{}{z} \right)P_1 + \frac{1}{\rho_0^2} \DD{\rho_0}{z} 
 \DP{P_1}{z} \nonumber \\
&& + 2 i C \Omega_K k'_x \frac{\bar{v}_0} {\rho_0} \rho_1
+ 2 i C \Omega_K  k'_x  v_1  -i \left( 2 C - 3T \right)\Omega_K k_y  u_1 
\nonumber \\
&& + \DP{}{z} \left( \frac{1}{\rho_0^2}\DP{P_0}{z} \rho_1 \right)
- 4\pi G \rho_1.
\label{eqn:div2}
\end{eqnarray}
We use the first order approximation:
\begin{equation}	
	\DP{D}{t} \approx \frac{D^{n+1} -D^{n}}{\Delta t}.
	\label{eqn:eul}
\end{equation}
We require that the equation of continuity (\ref{eqn:pt1}) is satisfied
in the next step, {\it i.e.} $D^{n+1}=0$. Thus, we get the equation for $P_1$:
\begin{eqnarray}
&& \left( -k'^{n2}_x -k^2_y  +  \DPt{}{z}  -
\frac{1}{\rho_0} \DD{\rho_0}{z}  \DP{}{z}   \right)P_1^n \nonumber \\	
&=& \rho_0 \left\{ \left(\frac{1}{\Delta t} - i k_y \bar{v}_0 \right) D^n
 - 2 i k_y \DD{\bar{v}_0}{z} w_1^n   + 2 i \Omega_K C  k'^n_x  v_1^n
  -i \left(2C -3T\right)\Omega_K  k_y  u_1^n \right.
\nonumber \\	
&& \left. + 2 i C \Omega_K   k'^n_x \frac{\bar{v}_0}{\rho_0} \rho_1^n 
+ \DP{}{z}\left(\frac{1}{\rho_0^2}\DP{P_0}{z} 
\rho_1^n \right) 
-4 \pi G \rho_1^n
\right\}.
\label{eqn:poisson}
\end{eqnarray}
From this equation, we can get $P_1$ at the time step $n$. 
We calculate $\Psi$ at the time step $n$ from 
\begin{equation}
 	\DPt{\Psi_1^n}{z}   - (k'^{n2}_x + k^{2}_y)  \Psi_1^n 
=  4 \pi G \rho_1^n.
\end{equation}

Next, we calculate
$\rho_1,u_1,v_1$ and $w_1$ at the time step $n+1$ from equations (\ref{eqn:pt2}) to (\ref{eqn:pt5}) using approximation like (\ref{eqn:eul}):
\begin{equation}
	 \rho_1^{n+1} =  \rho_1^{n}  
+ \Delta t \left\{-  i k_y \bar{v}_0 \rho_1^n
 - \DD{\rho_0} {z}w_1^{n} \right\},
	\label{eqn:spt2}
\end{equation}
\begin{equation}
	   u_1^{n+1} =  u_1^{n}  + \Delta t \left\{ -  i k_y \bar{v}_0  u_1^n
       - i k'^n_x \frac{1 }{\rho_0} P_1^n
    + 2 C\Omega_K \frac{\bar{v}_0}{\rho_0}\rho_1^n
	+ 2 C\Omega_{K}v_1^n  - i k'^n_x \Psi_1^n \right\},
	\label{eqn:spt3}
\end{equation}
\begin{equation}
       v_1^{n+1} =  v_1^{n}  + \Delta t \left\{ -i k_y \bar{v}_0 v_1^n
	- \DD{\bar{v}_0}{z} w_1^n
       - i k_y \frac{1}{\rho_0} P_1^n
      - \left(2 C- \frac{3}{2}T\right) \Omega_{K}u_1^n - i k_y \Psi_1^n 
 \right\},
	\label{eqn:spt4}
\end{equation}
\begin{equation}
 w_1^{n+1} =  w_1^{n}  + \Delta t \left\{ - i k_y \bar{v}_0  w_1^n
   - \frac{ 1}{\rho_0} \DP{P_1^n}{z}
  + \frac{1}{\rho_0^2}
  \DP{P_0}{z} 
  \rho_1^n -  \DP{\Psi_1^n}{z}\right\}.
	\label{eqn:spt5}
\end{equation}
The above method is the first order accuracy with respect to $\Delta t$.
In order to improve them to the second order accuracy, we use the following strategy. 
We replace perturbed quantities $\mbox{\boldmath{$f$}}_1^n \equiv (\rho_1^n, u_1^n, v_1^n, w_1^n)$ on right hand side of equation (\ref{eqn:poisson}) with 
\begin{equation}
 \mbox{\boldmath{$f$}}_1^{n+ \frac{1}{2} }   
= (\mbox{\boldmath{$f$}}_1^{n+1} + \mbox{\boldmath{$f$}}_1^{n}) /2.
\end{equation}
Again, we solve equation (\ref{eqn:poisson}) replacing $n$ 
by $n+ 1/2$.
More exact values of $\rho_1^{n+1},u_1^{n+1},v_1^{n+1}$ and $w_1^{n+1}$ are obtained by replacing these  quantities at $n$ in braces of the right hand sides 
of equations (\ref{eqn:spt2}) to (\ref{eqn:spt5}) with ones at $n+1/2$.
We adopt one dimension staggered mesh, where $w_1$ is 
estimated at grids and $P_1, \Psi_1, \rho_1,u_1,v_1$  at midpoints of adjacent meshes.  
In equations (\ref{eqn:poisson}), (\ref{eqn:spt2}) and (\ref{eqn:spt4}),
$w_1$ is calculated by taking the mean values at the adjacent meshes.
So as $\rho_1$ in equation (\ref{eqn:spt5}).

Initial conditions are set as follows.
Some Fourier components of lower orders are selected as to satisfy 
the boundary conditions of $u_1, w_1$ and $\rho_1$.
Each Fourier coefficient is given by a random number.
Velocity $v_1$ is determined from $u_1$ and $w_1$ by using the equation of  continuity (\ref{eqn:pt1}).




% \clearpage


\begin{figure}
\plotone{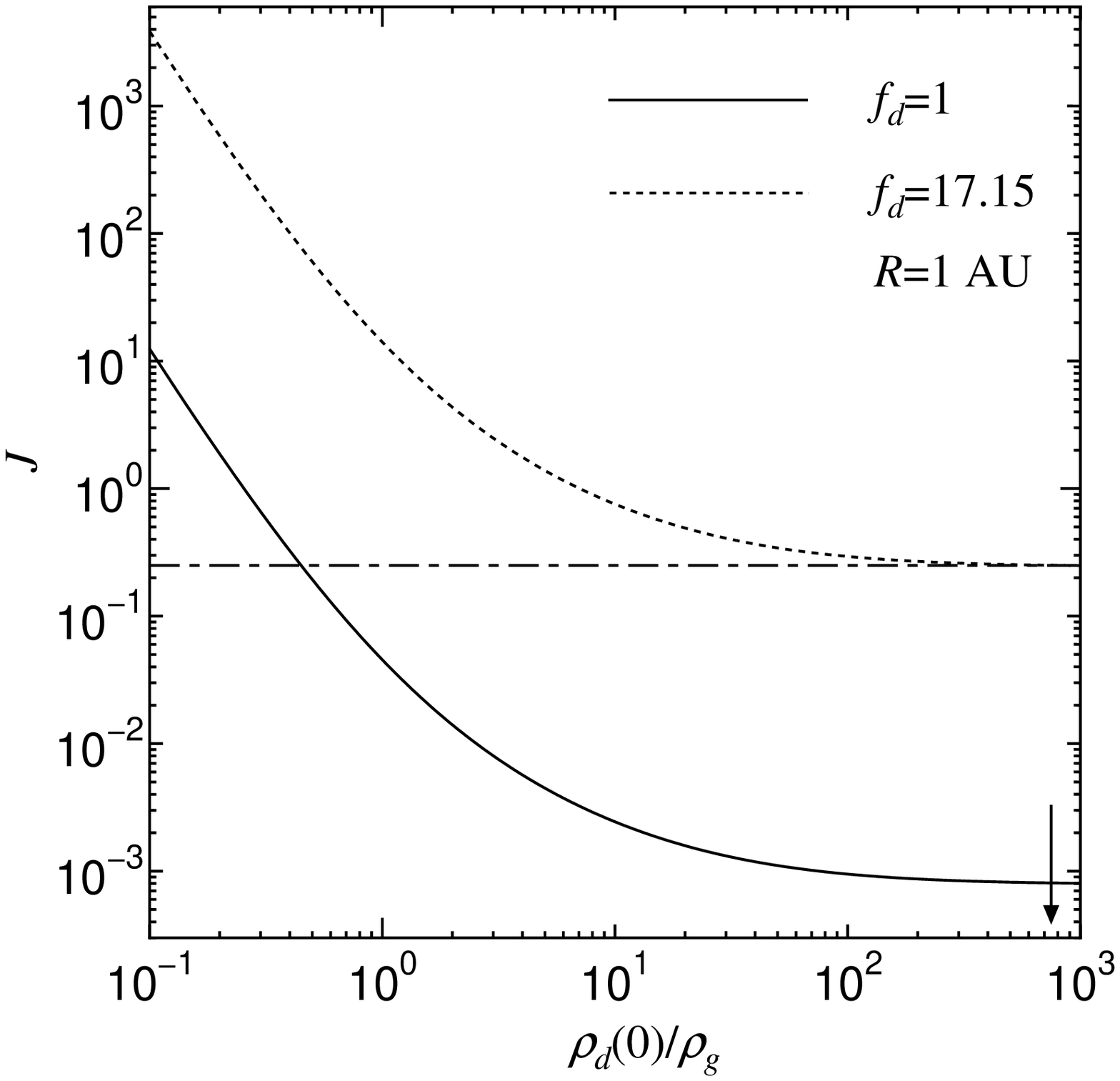}
\caption{ The Richardson number $J$ as 
functions of the dust/gas density ratio 
on the midplane $\rho_{d}(0)/\rho_g$
for $f_d=1$ (sold line) and 
$f_d=17.15$ 
(dotted line) at $R=$1AU. 
The  position of the arrow  shows the critical dust density of 
the gravitational instability.
The dot-dashed line denotes the 
critical Richardson number $J_c=0.25$. 
\label{fig1}}
\end{figure}

\begin{figure}
\plotone{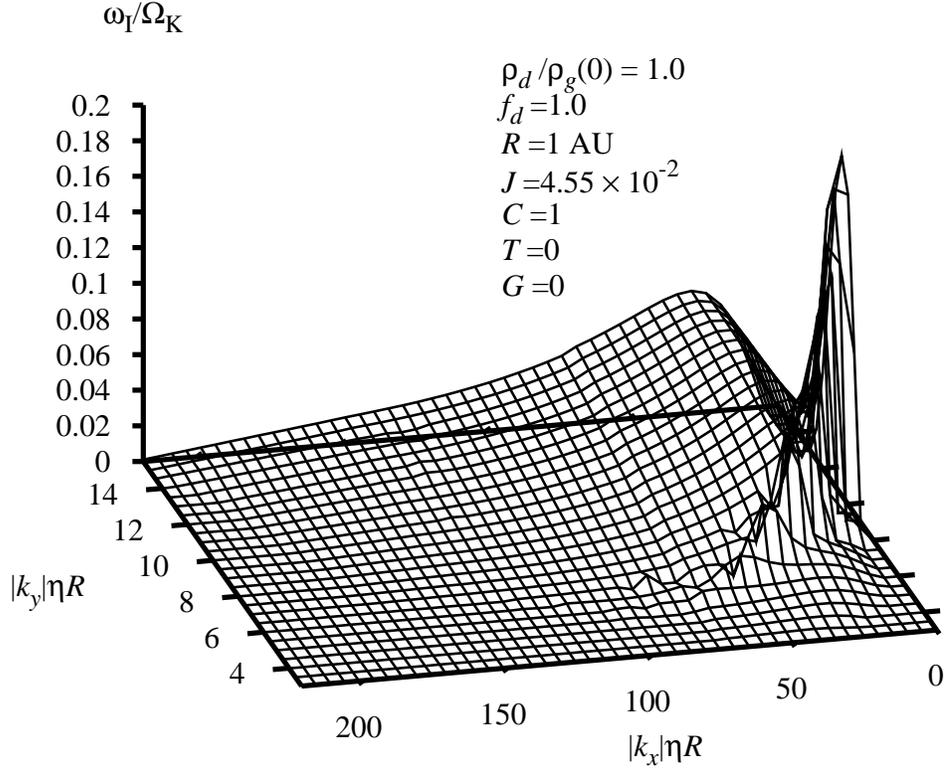}
\caption{ 
The growth rate $\omega_I$ of the mode 
as a function of the absolute values 
of the radial and the azimuthal wave numbers $|k_x|$ and $|k_y|$,
respectively, in the case where $C=1$, $T=0$ and $G=0$
with $J=4.55 \times 10^{-2}$, $\rho_{d}(0)/\rho_g=1.0$ 
and $f_d=1.0$ at $R=$1AU.
 \label{fig2}}
\end{figure}

\begin{figure}
\plotone{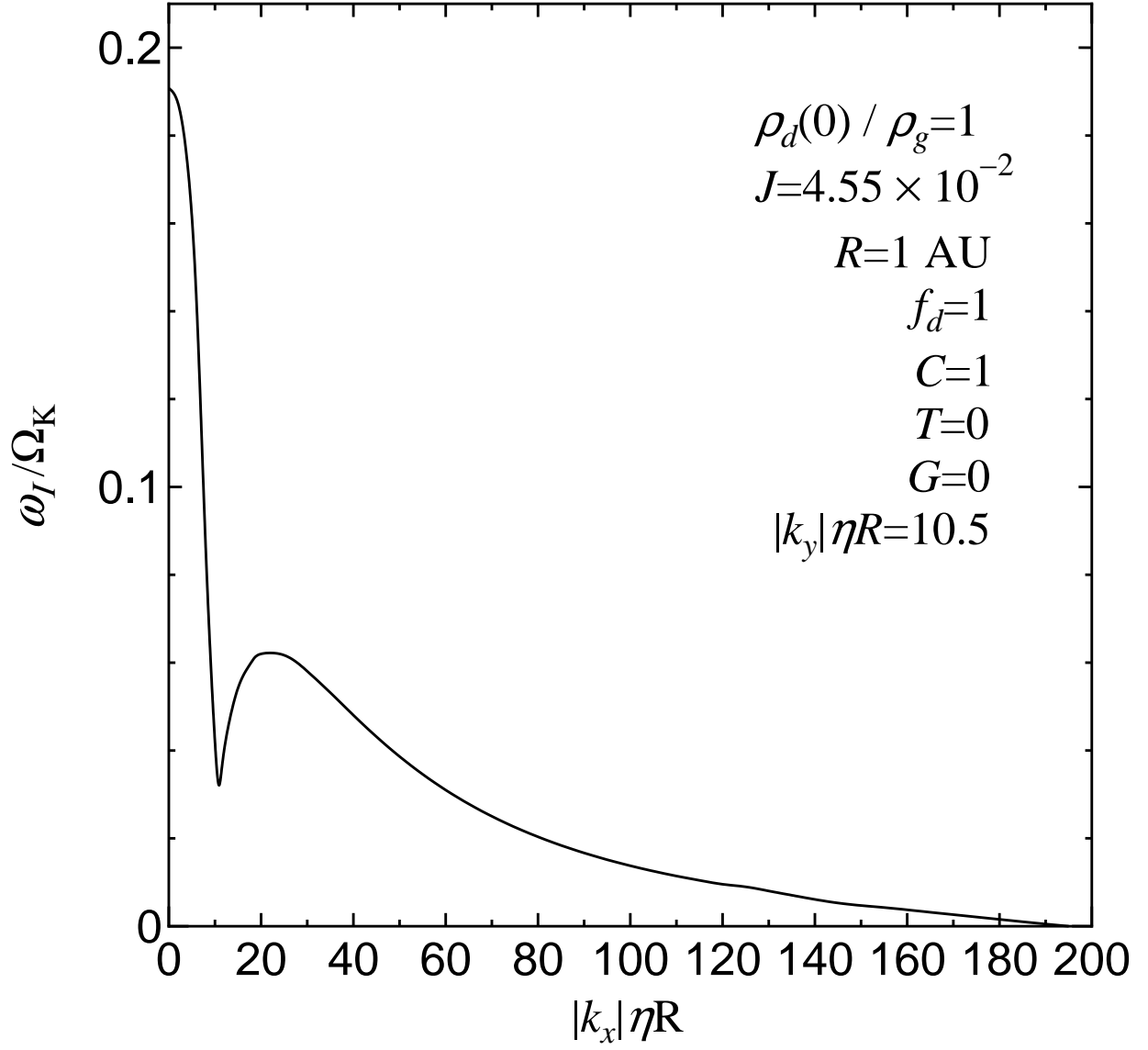}
\caption{ 
The growth rate $\omega_I$  as functions of 
$|k_x|$ in the case where $|k_y| \eta R=10.5$ 
with  $J=4.55 \times 10^{-2}$, 
$\rho_{d}(0)/\rho_g=1$, $C=1$, $T=0$ and $G=0$. 
 \label{fig3}}
\end{figure}

\begin{figure}
\plotone{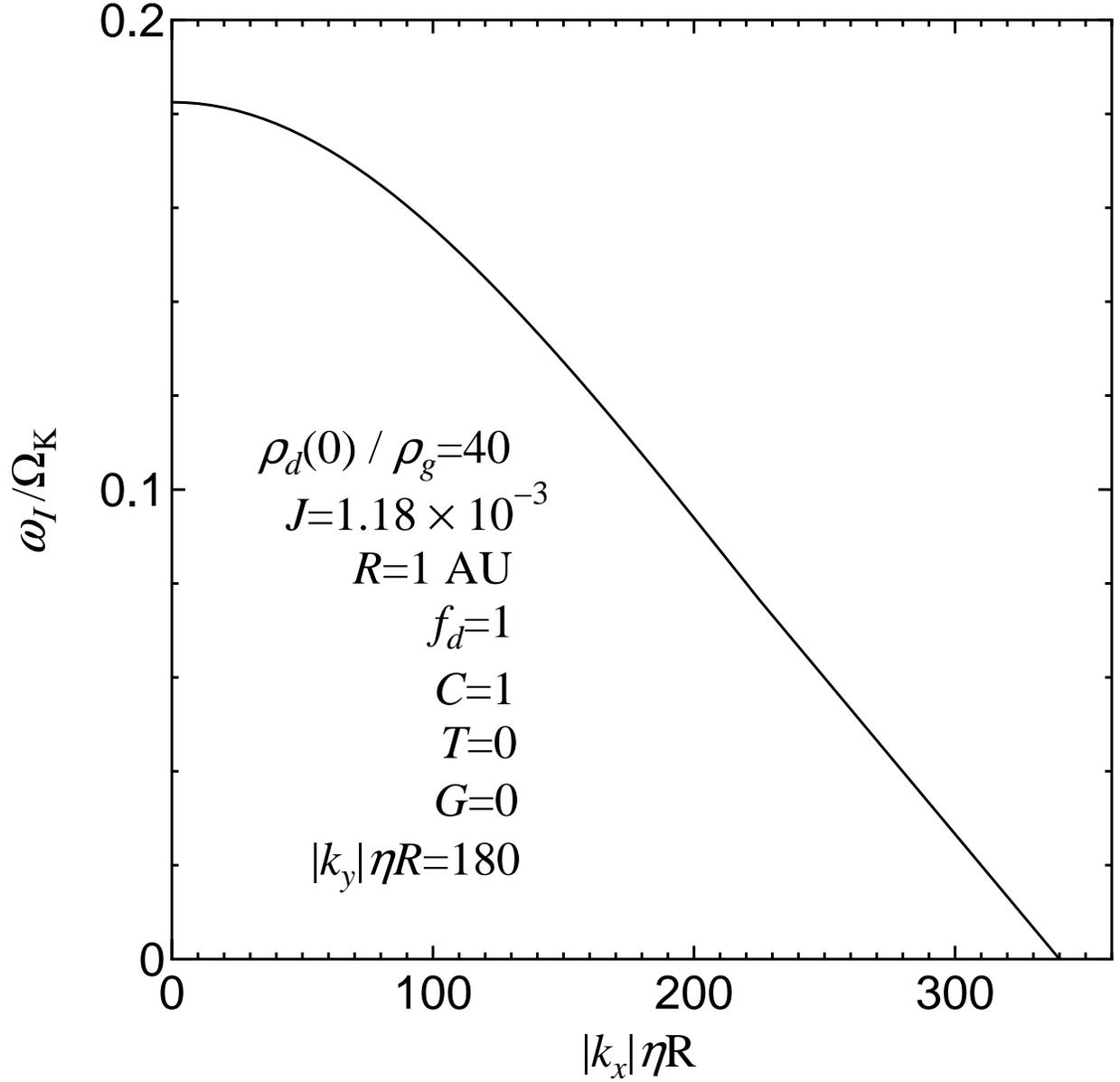}
\caption{ 
The growth rate $\omega_I$  as functions of 
$|k_x|$ in the case where $|k_y| \eta R=180$ 
with  $J=1.18 \times 10^{-3}$, 
$\rho_{d}(0)/\rho_g=40$, 
$C=1$, $T=0$ and $G=0$.
 \label{fig4}}
\end{figure}

\begin{figure}
\plotone{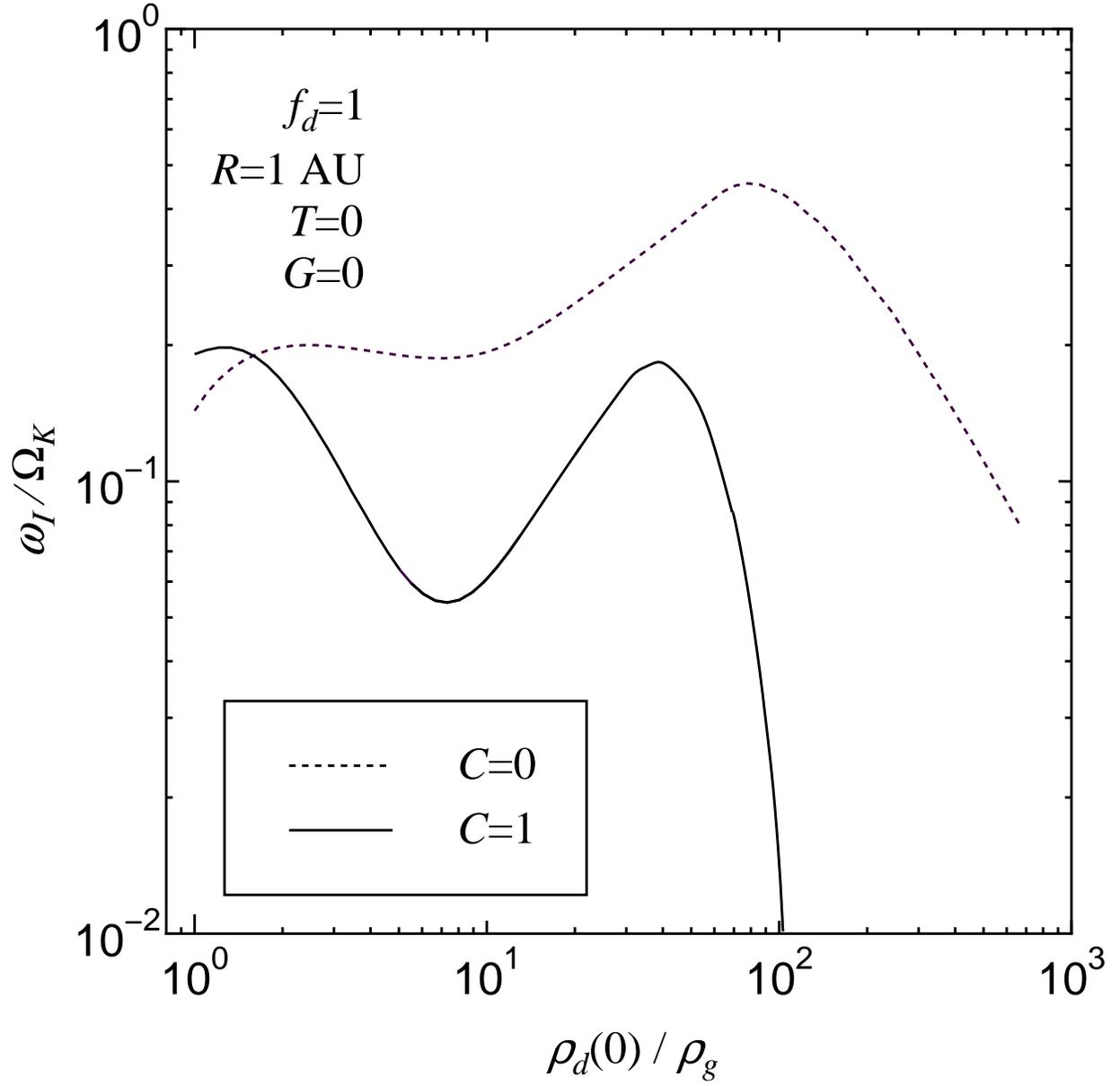}
\caption{ 
The growth rate $\omega_I$ of the mode with the most unstable 
wave number as 
functions of the 
dust/gas density ratio
on the midplane
$\rho_{d}(0)/\rho_g$, with 
the Coriolis force (solid curve) and
without this force (dotted curve). \label{fig5}}
\end{figure}

\begin{figure}
\plotone{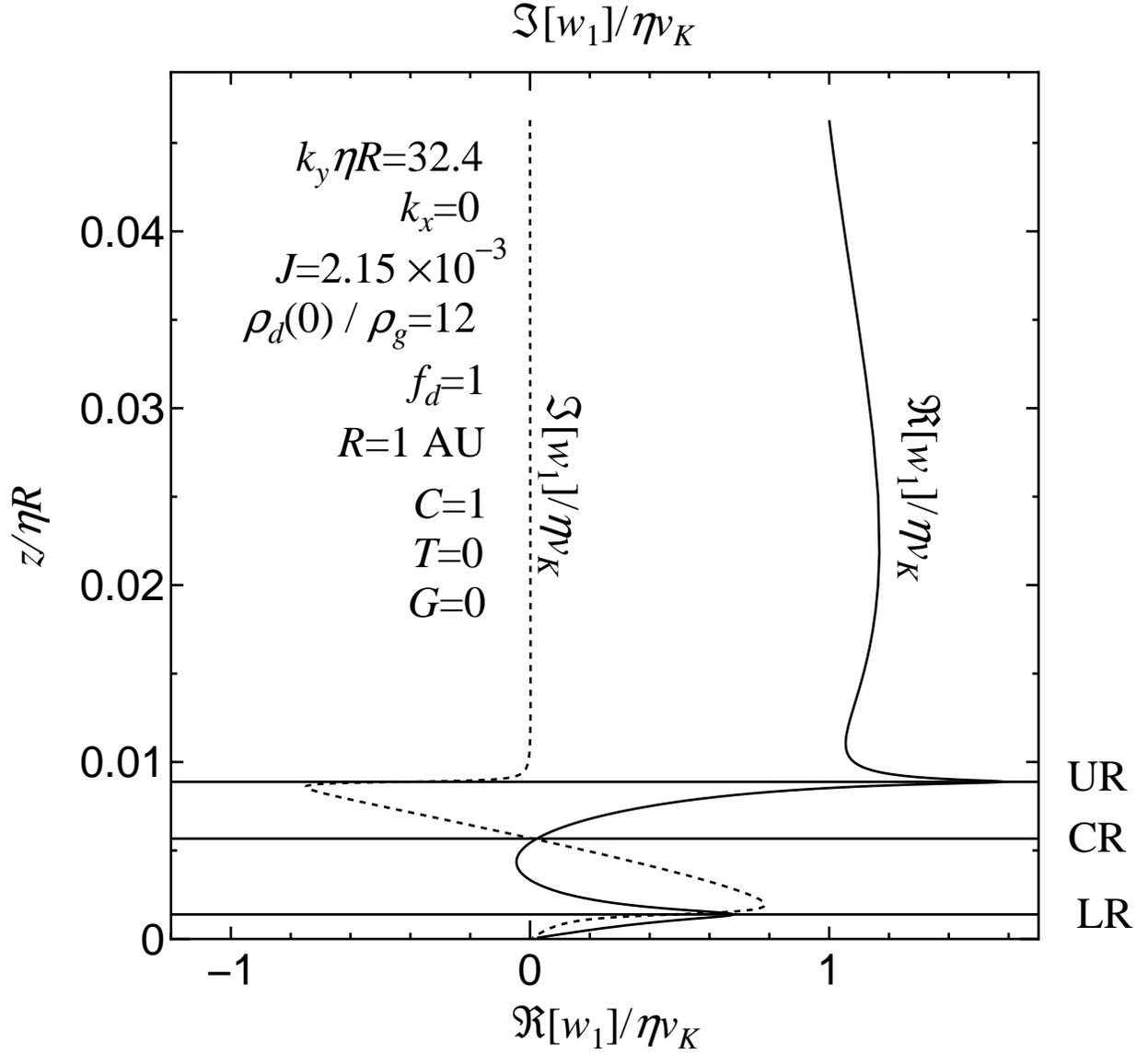}
\caption{ 
The real and the 
imaginary parts of the eigenfunction $\Re(w_1)$ and $\Im(w_1)$,
respectively, in the case where
$k_y \eta R=32.4$, $k_x =0$, $\rho_{d}(0)/\rho_g=12$, 
$C=1$, $T=0$, $G=0$ and $f_d=1$ at $R=$1AU. 
The horizontal line denoted by CR shows the co-rotation sheet.
The horizontal lines  denoted by UR and LR show the
upper and lower resonances, respectively.
 \label{fig6}}
\end{figure}

\begin{figure}
\plotone{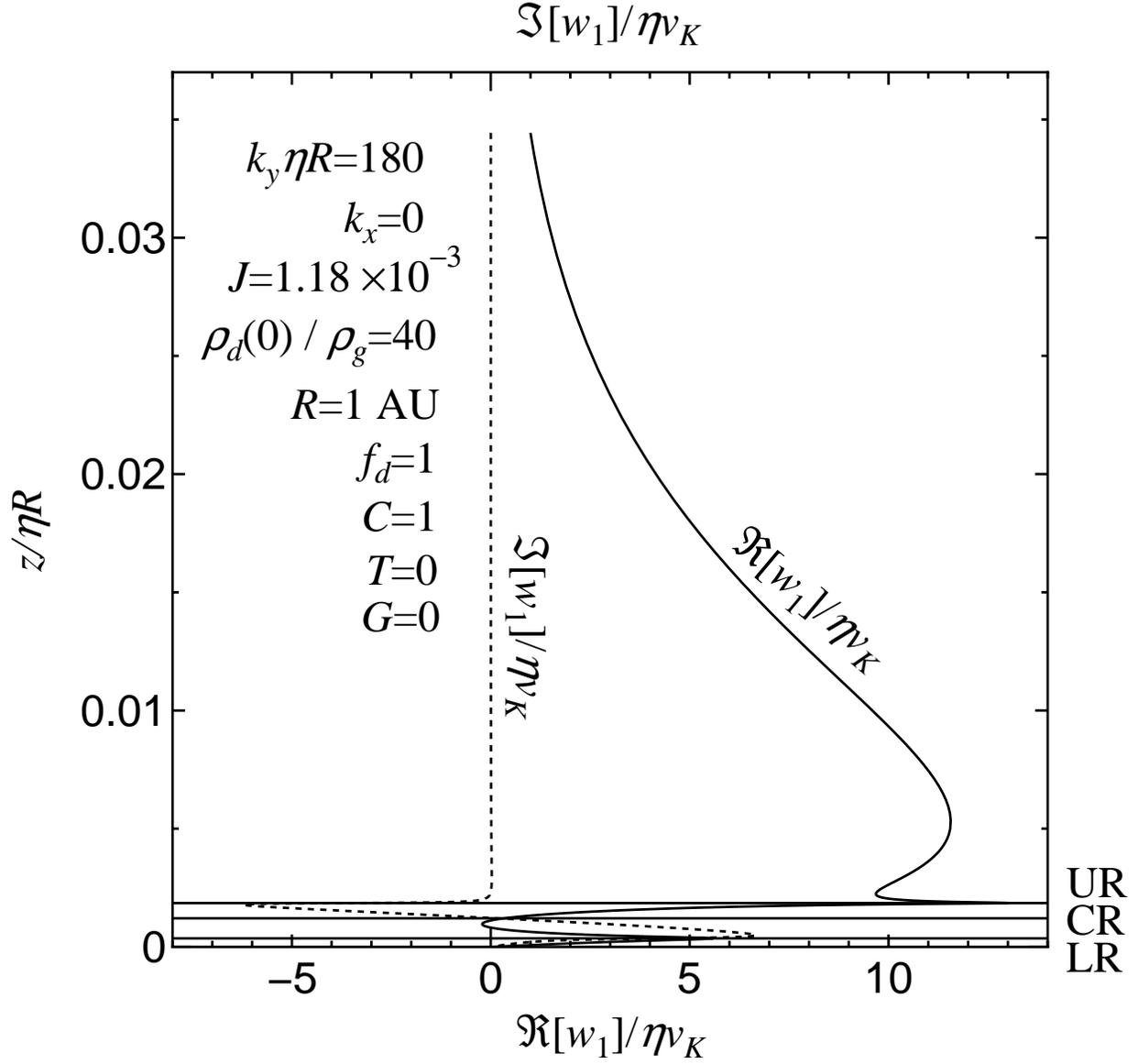}
\caption{ 
The real and imaginary parts of the eigenfunction $\Re(w_1)$ and $\Im(w_1)$,
respectively, in the case where 
$k_y \eta R=180$, 
$k_x =0$, $\rho_{d}(0)/\rho_g=40$, 
$C=1$, $T=0$, $G=0$ and $f_d=1$ at $R=$1AU. 
 \label{fig7}}
\end{figure}

\begin{figure}
\plotone{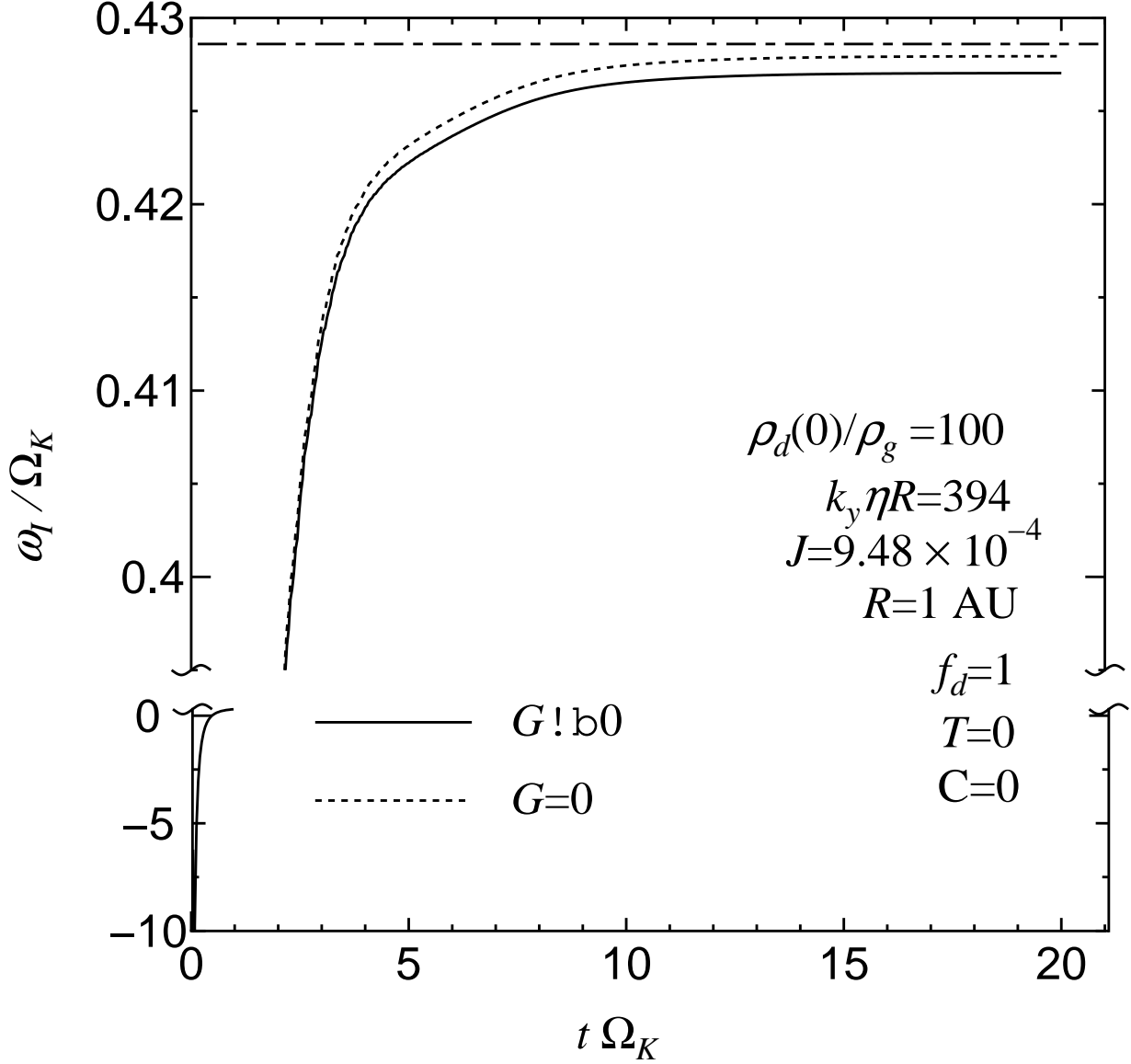}
\caption{ 
The time evolution of 
the growth rate in the case including ($G\ne 0$, sold line) and neglecting ($G=0$, dotted line) the self-gravity 
with  $J=9.48 \times 10^{-4}$, 
$\rho_{d}(0)/\rho_g=100$, $C=0$, $T=0$, and
$k_y \eta R=394$. 
The dotted line shows the 
solution derived by the method in section 4.
 \label{fig8}}
\end{figure}

\begin{figure}
\plotone{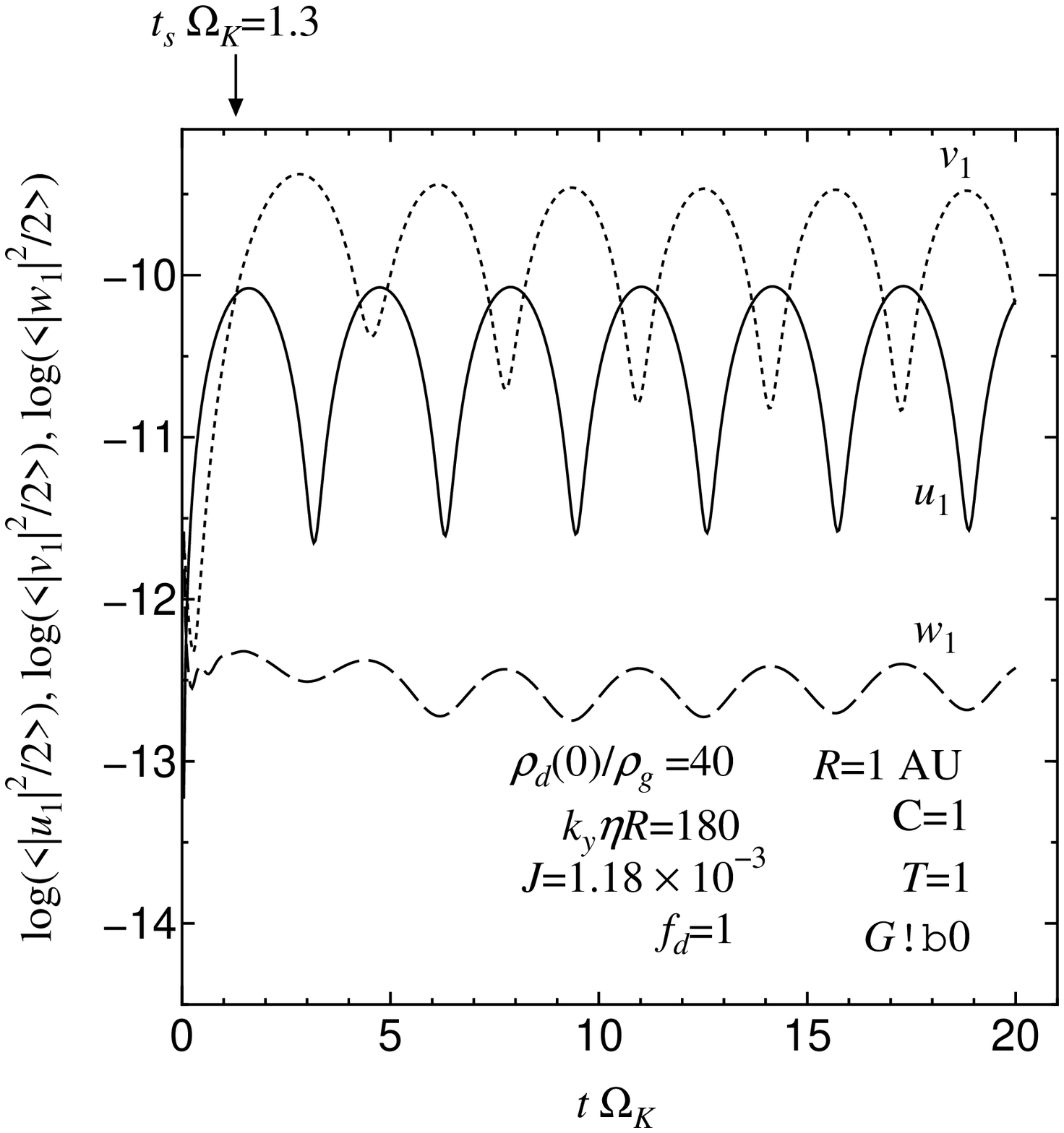}
\caption{ 
The time evolution of radial, azimuthal and vertical
components
of the 
the spatially averaged kinetic energy of the perturbed flow, 
in the case
including the 
tidal force ($T=1$)
with  $J=1.17 \times 10^{-3}$, 
$\rho_{d}(0)/\rho_g=40$, $C=1$, 
$G\ne 0$, $k_x=0$ and $k_y \eta R=180$ at $R=$1AU. 
 \label{fig9}}
\end{figure}

\begin{figure}
\plotone{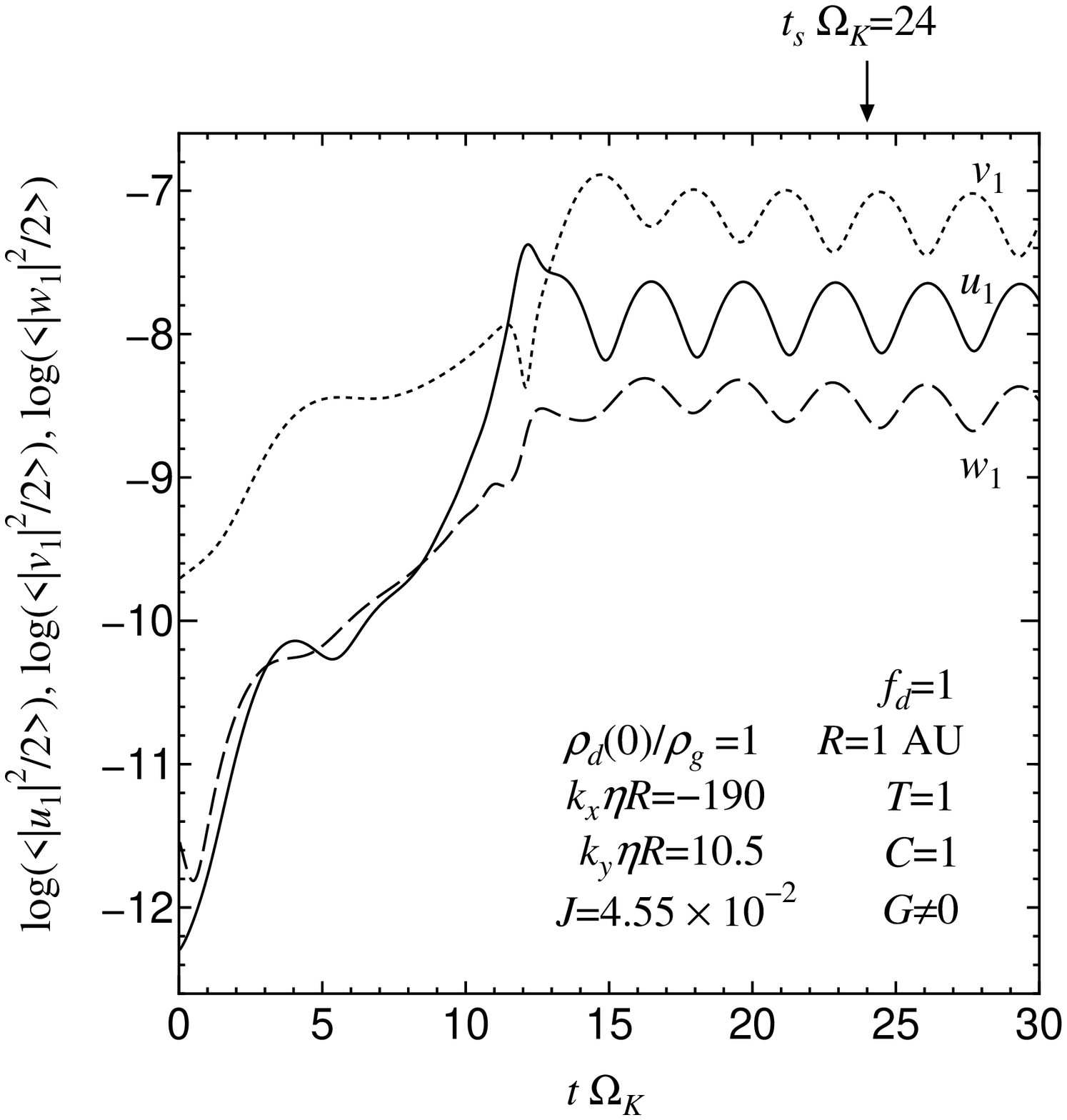}
\caption{ 
The time evolution of radial, azimuthal and vertical
components
of the 
the spatially averaged kinetic energy of the perturbed flow, 
in the case
including the
tidal force ($T=1$) 
with  $J=4.55 \times 10^{-2}$, 
$\rho_{d}(0)/\rho_g=1$, $C=1$,
$G\ne 0$, $k_x \eta R=-190$ and $k_y \eta R=10.5$ at $R=$1AU. 
 \label{fig10} }
\end{figure}

\begin{figure}
\plotone{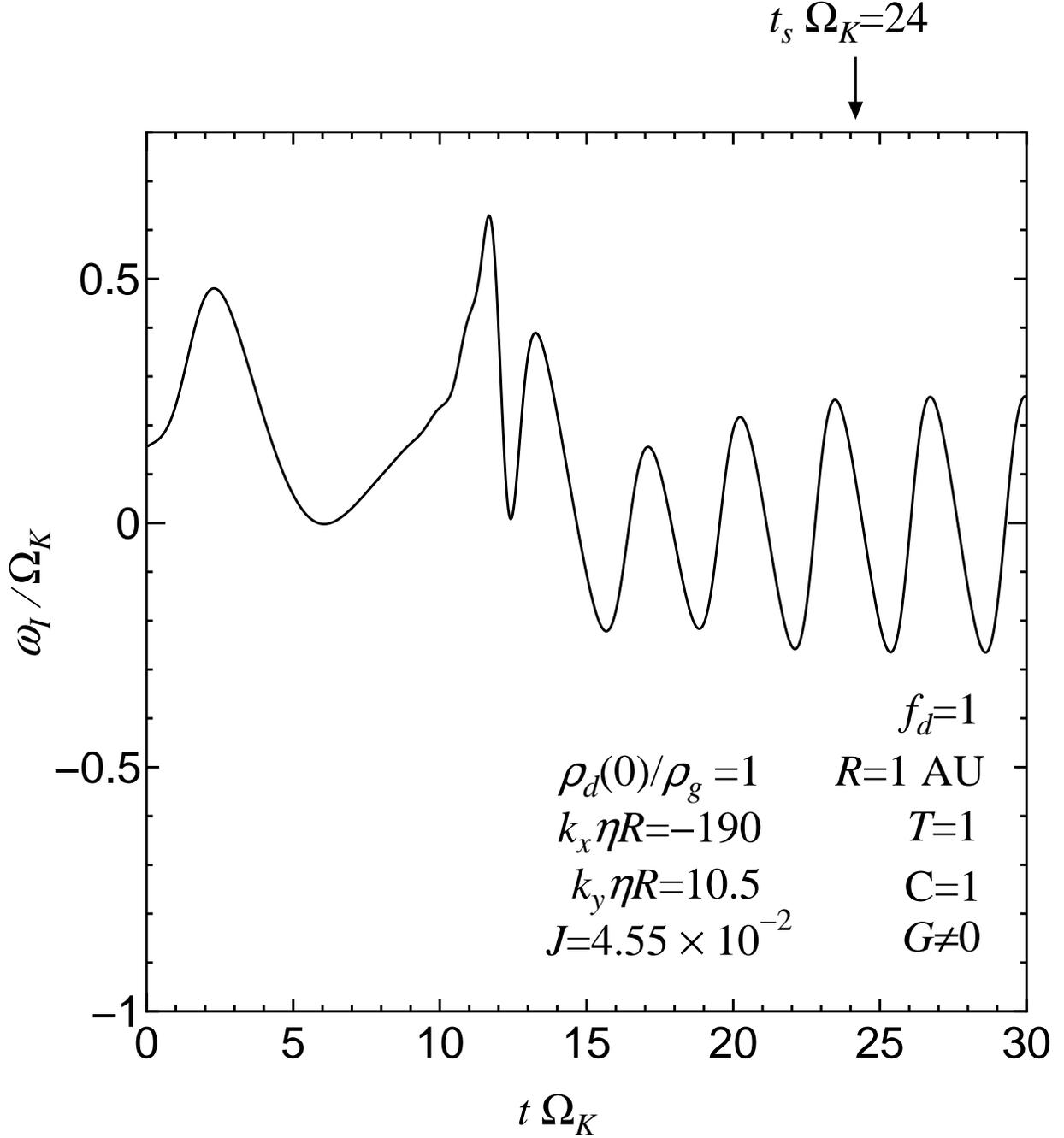}
\caption{ 
Time evolution of  the growth rate 
in the case
including 
tidal force ($T=1$) 
with  $J=4.55 \times 10^{-2}$, 
$\rho_{d}(0)/\rho_g=1$, $C=1$, 
$G\ne 0$, $k_x \eta R=-190$ and $k_y \eta R=10.5$ at $R=$1AU. 
 \label{fig11}}
\end{figure}
%\clearpage 

\end{document}